# 2D-to-3D transformation of ring origami via snap-folding instabilities

Lu Lu, Sophie Leanza, Luyuan Ning, Ruike Renee Zhao*

Department of Mechanical Engineering, Stanford University, Stanford, CA 94305, United States

* Corresponding author: rrzhao@stanford.edu (R.R.Z.)

## Abstract

Shape-morphing structures have been intensively researched for a wide range of engineering applications, including deployable emergency shelters, multimodal soft robots, and property-programmable metamaterials, owing to their ability to change shape and size in response to external stimuli. Ring origami, consisting of closed-loop rods, is a class of shape-morphing structures that undergo shape transformation through folding enabled by snap-buckling instabilities, referred to as snap-folding instabilities. Previous studies have shown that 2D ring origami composed of rod segments with in-plane natural curvature (i.e., the stress-free curved state lies in the plane of the planar ring) can achieve diverse and intriguing 2D-to-2D shape transformations. Here, we propose a 2D-to-3D shape transformation strategy for ring origami by introducing out-of-plane natural curvature (i.e., the stress-free curved state lies in a plane perpendicular to the planar ring) into the rod segments. Due to natural curvature-induced out-of-plane bending moments, a 2D elastic ring spontaneously snaps out of plane and reaches equilibrium in a 3D configuration. These snapping-induced out-of-plane shape transitions not only enable self-guided, spontaneous shape morphing, but also allow the construction of complex structures from simple geometries, making them promising for the design of functional deployable and foldable structures. By combining a multi-segment Kirchhoff rod model with finite element simulations and experiments, we systematically investigate the 3D equilibrium states and transition behavior of these systems. Using square and hexagonal rings as representative examples, we demonstrate that by rationally designing the out-of-plane natural curvature of rod segments, 2D rings can exhibit a range of functional behaviors, including spontaneous 2D-to-3D shape transformation (e.g., planar square to sphere) via snap-folding, multistability with various 3D configurations, and monostability with a compact zero-energy 3D configuration. Our work presents a new design strategy that enables complex shape-morphing structures from simple geometries and realizes multistability and spontaneous 2D-to-3D transformation via snap-folding instabilities, which holds great potential in applications such as deployable aerospace structures, soft robotics, and mechanical metamaterials.

*Keywords: Ring origami; Snap-folding; Shape transformation; Multistability; Natural curvature.*

## 1. Introduction

Shape-morphing structures (Yang et al., 2023b) have gained significant attention for their broad engineering applications, including deployable emergency shelters (Melancon et al., 2021; Thrall and Quaglia, 2014), multimodal soft robots (Chen et al., 2022; Wu et al., 2021b; Xu et al., 2025), and property-programmable metamaterials (Chen et al., 2021; Hwang et al., 2022; Liu et al., 2024; Meng et al., 2023), owing to their ability to change shape and size in response to external stimuli. Through careful structural design, these structures can achieve shape transformations of various forms, such as 1D-to-2D (Rahman et al., 2024; Schenk and Guest, 2013), 2D-to-2D (Chen et al., 2019; Rafsanjani and Pasini, 2016), 2D-to-3D (Dang et al., 2022; Liu et al., 2023; Yang et al., 2023a; Zhang et al., 2022a; Zirbel et al., 2013), and 3D-to-3D (Dang et al., 2025; Hoberman, 1991; Ma et al., 2020; Riccobelli et al., 2023). Among these, 2D-to-3D shape transformations are particularly attractive due to the ease of manufacturing the initial planar configuration and the design flexibility of the resulting 3D structures, which can be tailored to meet specific functional or environmental requirements. This strategy has been widely applied in areas such as flexible electronics (Cheng et al., 2023; Fan et al., 2020; Jin et al., 2023) and morphable biomedical devices (An et al., 2024b; Zhang et al., 2021). Moreover, in certain 2D-to-3D shape transformations, the inherent contrast between the large surface area of the flat 2D form and the compact volume of its folded 3D form is especially advantageous for large-scale deployable aerospace applications, such as solar arrays (Chen et al., 2019; Pehrson et al., 2020; Zirbel et al., 2013) and solar sails (Block et al., 2011; Koryo, 1985), where compact, efficient storage and transport are critical prior to deployment. Origami and kirigami principles (i.e., the art of paper folding and cutting) (Callens and Zadpoor, 2018; Jin and Yang, 2024; Lu et al., 2023c; Misseroni et al., 2024), as well as elastic instabilities such as buckling and wrinkling (An et al., 2024a; Bo et al., 2023; Guo et al., 2025; Shuai et al., 2023; Siéfert et al., 2019; Xu et al., 2015), have been widely utilized for 2D-to-3D shape transitions. In these systems, a flat 2D structure can morph into diverse 3D configurations through carefully engineered 2D geometries and external mechanical loads or stimuli. However, these approaches mostly rely on continuous and precisely controlled external mechanical constraints to maintain the desired 3D shape, which increases system complexity and limits their practical implementation.

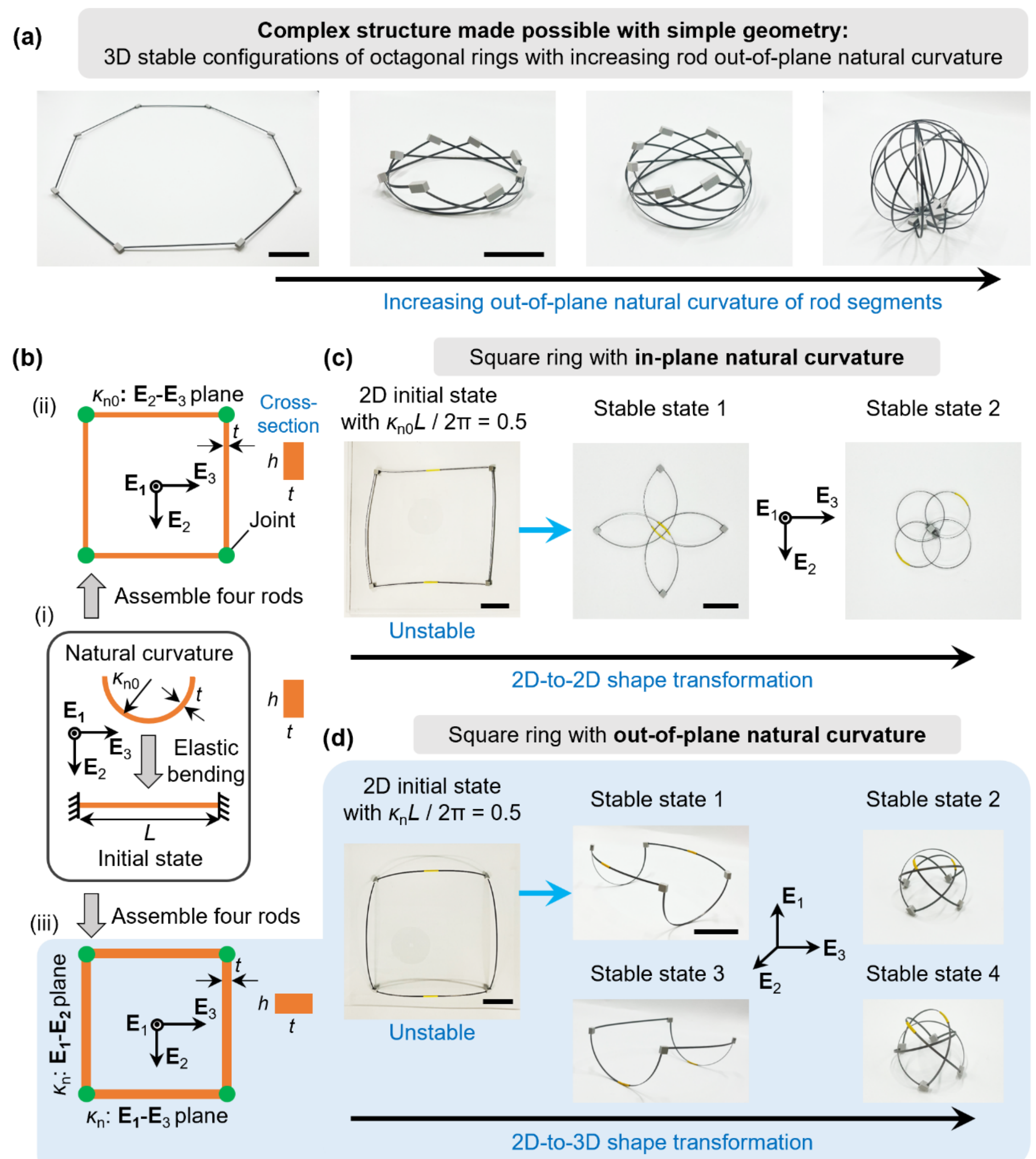

**Fig. 1.** Ring origami consisting of elastic rod segments. (a) Complex structure made possible with simple geometry: 3D stable configurations of stainless-steel octagonal rings obtained by increasing the out-of-plane natural curvature of the rod segments. The leftmost octagonal ring is planar and composed of rod segments with zero natural curvature. The initial planar states of the three 3D configurations are unstable

and can spontaneously snap into these stable configurations. (b-i) Schematic of an elastic rod of length $L$, cross-sectional width $t$ and height $h$, with natural curvature $\kappa_{n0}$ in the stress-free state. The naturally curved rod can be elastically straightened via bending and remain stable when both ends are clamped. (b-ii) Four identical such rod segments can be assembled into a square ring through rigid joints, in which the natural curvature of the rod segments lies in the $\mathbf{E}_2$-$\mathbf{E}_3$ plane of the planar ring, resulting in a square ring with in-plane natural curvature $\kappa_{n0}$. (b-iii) The same four rod segments can be alternatively assembled into another square ring, in which the natural curvature of the rod segments lies either in the $\mathbf{E}_1$-$\mathbf{E}_3$ or $\mathbf{E}_1$-$\mathbf{E}_2$ plane that is perpendicular to the $\mathbf{E}_2$-$\mathbf{E}_3$ plane of the planar ring, resulting in a square ring with out-of-plane natural curvature $\kappa_n$. (c) 2D-to-2D shape transformation of a stainless-steel square ring with dimensionless in-plane natural curvature $\kappa_{n0}L/2\pi = 0.5$. (d) 2D-to-3D shape transformation of a stainless-steel square ring with dimensionless out-of-plane natural curvature $\kappa_n L/2\pi = 0.5$. In both (c) and (d), the 2D initial state is unstable and maintained in the plane by a transparent plate. Once the constraint is removed, the ring spontaneously snaps into a stable state, which can further transition into other stable states under external stimuli such as bending. In (d), stable states 1 and 3, as well as stable states 2 and 4, share the same configuration. However, the same rod segments in the two states undergo different deformations or stack in different ways. In the dome-like configurations, the edges marked with yellow tape lie beneath the other pair of edges in stable state 2 but lie above them in stable state 4. Scale bars: 5 cm. The folded states in (a), (c), and (d) share the same scale bar as the one indicated.

In this work, we propose a simple yet effective design strategy to realize 2D-to-3D shape transformations using ring origami systems consisting of 2D closed-loop segmented rods with out-of-plane natural curvature (i.e., the stress-free curved state lies in a plane perpendicular to the planar ring). The proposed strategy not only enables self-guided, spontaneous shape morphing by harnessing snap-folding instabilities, but also allows the construction of complex structures from simple geometries. For example, **Fig. 1(a)** illustrates several distinct 3D stable configurations obtained from increasing the out-of-plane natural curvature of a stainless-steel octagonal ring's rod segments. Under a certain natural curvature, the ring spontaneously snaps from its unstable initial planar configuration to a complex 3D stable configuration (experimental demonstrations are presented in **Video 1** in the **Supplementary Material**, and experimental details on the ring fabrication are provided in **Appendix A**). Prior studies have demonstrated that ring origami composed of closed-loop rods with in-plane natural curvature (i.e., the stress-free curved state lies in the plane of the planar ring) exhibit diverse and intriguing 2D-to-2D shape transformations (Audoly and Seffen, 2015; Goto et al., 1992; Lu et al., 2023a; Lu et al., 2024; Sun et al., 2022; Wu et al., 2022; Wu et al., 2021a). Here, we show that introducing out-of-plane natural curvature to the rod significantly expands the design space, enabling enhanced 3D shape programmability and multistability by the spontaneous snap-folding of 2D planar rings.

To illustrate the design strategy, **Fig. 1(b)** compares the ring origami constructed by the same rod segments with either in-plane or out-of-plane natural curvature. A square ring is used as a representative example. As shown in **Fig. 1(b-i)**, consider an elastic rod of length $L$, cross-sectional width $t$ and height $h$, with natural curvature $\kappa_{n0}$ in its stress-free state. Through bending, the rod can be elastically straightened and maintained in a straight configuration under external constraints. Four such rod segments can be assembled into a square ring, in which the natural curvature of each rod segment lies in the plane of the planar ring (the $\mathbf{E}_2$-$\mathbf{E}_3$ plane), resulting in a 2D ring with in-plane natural curvature $\kappa_{n0}$ (**Fig. 1(b-ii)**). Within a certain range of natural curvature, such rings with properly designed cross-sectional aspect ratios can exhibit multiple planar equilibrium configurations (Audoly and Seffen, 2015; Leanza et al., 2024b; Lu et al., 2023b; Lu et al., 2024; Manning and Hoffman, 2001). For example, as shown in **Fig. 1(c)**, for a square ring composed of four identical stainless-steel rod segments with cross-sectional aspect ratio $h/t$ = 4, when the dimensionless in-plane natural curvature, $\kappa_{n0}L/2\pi$, equals 0.5, the ring becomes unstable (stable with $\kappa_{n0}L/2\pi$ up to 0.4, and a positive in-plane natural curvature induces a moment that bends the rod segment toward the ring's center). Note that throughout the paper, the cross-sectional dimension of the rod in the plane of the planar ring is defined as width $t$, while the dimension perpendicular to this plane is defined as height $h$. Upon releasing the external constraints, the unstable ring spontaneously snaps into a 2D propeller-like configuration (stable state 1), which can transition into a 2D four-circle configuration (stable state 2) under external loads (see **Video 2** in the **Supplementary Material**), enabling 2D-to-2D shape transformations.

In contrast, the same four rod segments can be assembled into another square ring in which the natural curvature of each rod segment lies in a plane (either the $\mathbf{E}_1$-$\mathbf{E}_3$ or $\mathbf{E}_1$-$\mathbf{E}_2$ plane) that is perpendicular to the ring's planar configuration (the $\mathbf{E}_2$-$\mathbf{E}_3$ plane), resulting in a 2D ring with out-of-plane natural curvature $\kappa_n$ (**Fig. 1(b-iii)**). Since such a ring cannot remain stable in the plane due to the natural curvature-induced out-of-plane bending moment, it spontaneously snaps into a 3D configuration to reach equilibrium, enabling a 2D-to-3D shape transformation. As shown in **Fig. 1(d)**, a square ring composed of the same rod segments used in **Fig. 1(c)**, with dimensionless out-of-plane natural curvature $\kappa_n L/2\pi = 0.5$, exhibits four 3D stable states that can transition among one another (see **Video 2** in the **Supplementary Material**). Based on the previous definition, the cross-sectional aspect ratio $h/t$ in this case becomes 1/4, because the cross-section of each rod segment is rotated by 90° compared to that of the square ring with in-plane natural curvature.

Among the four stable states, two (states 1 and 3) share a deployed tent-like configuration, while the other two (states 2 and 4) share a folded dome-like configuration. Note that although the two stable states share the same configuration, the same rod segments in the two states undergo different deformations (states 1 and 3) or stack in different ways (states 2 and 4). As will be demonstrated in **Section 3**, by increasing the out-of-plane natural curvature of the rod segments, more complex 3D stable configurations can be obtained from the simple planar geometry.

To uncover the fundamental mechanics of the 2D-to-3D shape transformation, we systematically investigate the equilibrium states and transition behavior of 2D segmented rings with out-of-plane natural curvature through a combination of theoretical modeling, finite element simulations, and experiments. The theoretical framework is established based on the Kirchhoff rod theory (Audoly and Pomeau, 2010), which enables the prediction of 3D equilibrium states and transition behavior of 2D segmented rings with different out-of-plane natural curvatures. Using square and hexagonal rings as representative examples, we explore how the out-of-plane natural curvature affects the resulting 3D equilibrium states and transition behavior of 2D segmented rings. Two cases are discussed in detail: (i) square and hexagonal rings in which all rod segments have the same out-of-plane natural curvature, and (ii) square and hexagonal rings in which the rod segments have alternating positive and negative out-of-plane natural curvatures of equal magnitude (a positive out-of-plane natural curvature induces a moment that bends the rod segment upward, while a negative one induces a moment that bends it downward). The theoretical predictions are validated by finite element analysis (FEA) simulations and experiments, showing excellent agreements. We believe that the proposed strategy for constructing complex shape-morphing structures from simple geometries, combined with multistability and spontaneous 2D-to-3D shape transformation via self-guided snap-folding, could be promising in the design of functional applications, such as deployable aerospace structures (Leanza et al., 2022), soft robotics (Kim et al., 2023; Zhao et al., 2023), and mechanical metamaterials (Niloy et al., 2025).

The remainder of this paper is organized as follows. In **Section 2,** a multi-segment Kirchhoff rod model for predicting the equilibrium states and transition behavior of 2D segmented rings with out-of-plane natural curvature is introduced. In **Section 3**, we study the equilibrium states and transition behavior of 2D square and hexagonal rings with out-of-plane natural curvature. In **Section 4**, the main conclusions are summarized.

## 2. Theoretical model

We start by introducing the multi-segment rod model we used for the analysis of 2D segmented rings with out-of-plane natural curvature. The multi-segment rod model (Sun et al., 2022; Yu et al., 2021) is established based on the Kirchhoff rod theory (Audoly and Pomeau, 2010), which has been used for studying the folding behavior of stress-free polygonal rings (Sun et al., 2022), transition behavior of curved-sided hexagram rings with in-plane natural curvature (Lu et al., 2023a), static equilibria and bifurcations in bigons and bigon rings (Larsson and Adriaenssens, 2025; Yu et al., 2021), folding of closed-loop ribbons with creases or kinks (Huang et al., 2024; Yu et al., 2023), and bifurcation and multistability of serpentine strips (Shi et al., 2025). Here, we briefly outline the main governing equations, with details of the numerical implementation provided in **Appendix B**.

The elastic rings we study are composed of multiple rod segments of equal length $L$, cross-sectional width $t$, and height $h$, as illustrated in **Fig. 2(a)** using a square ring. Each rod segment is assumed to be unshearable and inextensible, and has a Young's modulus $E$ and Poisson's ratio $\nu$. The global material frame ($\mathbf{E}_1$, $\mathbf{E}_2$, $\mathbf{E}_3$) is introduced at the center of the ring, where $\mathbf{E}_2$ and $\mathbf{E}_3$ lie along two in-plane orthogonal directions, and $\mathbf{E}_1$ points out of the plane. The position vector $\mathbf{p}(s)$, which describes the spatial location of the centerline during deformation, is defined in the global material frame as $\mathbf{p}(s) = p_1(s)\mathbf{E}_1 + p_2(s)\mathbf{E}_2 + p_3(s)\mathbf{E}_3$, where $s$ denotes the arc length coordinate and $s \in [0, L]$. The local material frame [$\mathbf{e}_1(s)$, $\mathbf{e}_2(s)$, $\mathbf{e}_3(s)$] is attached to the rod's centerline, with $\mathbf{e}_1$ oriented along the height direction, $\mathbf{e}_2$ along the width direction, and $\mathbf{e}_3$ tangent to the centerline, which indicates $\mathbf{p}' = \mathbf{e}_3$. Throughout the paper, the prime $(\bullet)'$ denotes differentiation with respect to the arc length coordinate $s$. The associated curvatures in the three directions of the local material frame are denoted by $\kappa_1(s)$, $\kappa_2(s)$, and $\kappa_3(s)$, where $\kappa_1$ and $\kappa_2$ are the bending curvatures and $\kappa_3$ is the twisting curvature. The kinematics of the local material frame are governed by $\mathbf{e}_i' = \boldsymbol{\omega} \times \mathbf{e}_i$ for $i$ = 1, 2, 3, where $\boldsymbol{\omega} = \kappa_1\mathbf{e}_1 + \kappa_2\mathbf{e}_2 + \kappa_3\mathbf{e}_3$ is the Darboux vector. According to the Kirchhoff rod theory, the static equilibrium equations in the absence of any body forces and couples are given by

$$\mathbf{N}' = 0, \quad \mathbf{M}' + \mathbf{e}_3 \times \mathbf{N} = 0, \tag{1}$$

where $\mathbf{N}(s)$ and $\mathbf{M}(s)$ are internal forces and moments defined in the local material frame as $\mathbf{N} = N_1\mathbf{e}_1 + N_2\mathbf{e}_2 + N_3\mathbf{e}_3$ and $\mathbf{M} = M_1\mathbf{e}_1 + M_2\mathbf{e}_2 + M_3\mathbf{e}_3$. Moreover, linear constitutive relations are considered: $M_1 = EI_1(\kappa_1 - \kappa_0)$, $M_2 = EI_2(\kappa_2 - \kappa_\mathrm{n})$, and $M_3 = GJ\kappa_3$, where $\kappa_0$ is the initial in-plane curvature of the rod segment, and $\kappa_\mathrm{n}$ is the out-of-plane natural curvature. Moreover, $I_1$ and $I_2$ are the moments of inertia about the width and height directions, respectively, and $J$ is the rotational constant (Timoshenko and Goodier, 1951),

$$I_1 = \frac{1}{12}ht^3,\ \ I_2 = \frac{1}{12}h^3t,\ \ J = \frac{ht^3}{3}\left[1 - \frac{192}{\pi^5}\frac{t}{h}\sum_{k=1}^{\infty}\frac{1}{(2k-1)^5}\tanh\left(\frac{(2k-1)\pi h}{2t}\right)\right]. \tag{2}$$

Projecting Eq. (1) along $\mathbf{e}_1$, $\mathbf{e}_2$, and $\mathbf{e}_3$, and applying the constitutive relations yields

$$\begin{aligned}
&N_1' = N_2\kappa_3 - N_3\kappa_2,\\
&N_2' = N_3\kappa_1 - N_1\kappa_3,\\
&N_3' = N_1\kappa_2 - N_2\kappa_1,\\
&\kappa_1' = \kappa_0' + \frac{N_2}{EI_1} + \frac{\beta - 1}{\alpha}\kappa_2\kappa_3 - \frac{\beta}{\alpha}\kappa_\mathrm{n}\kappa_3,\\
&\kappa_2' = \kappa_\mathrm{n}' - \frac{N_1}{EI_2} - \frac{\alpha - 1}{\beta}\kappa_1\kappa_3 + \frac{\alpha}{\beta}\kappa_0\kappa_3,\\
&\kappa_3' = (\alpha - \beta)\kappa_1\kappa_2 - \alpha\kappa_0\kappa_2 + \beta\kappa_\mathrm{n}\kappa_1,
\end{aligned} \tag{3}$$

where $\alpha = EI_1 / GJ$ and $\beta = EI_2 / GJ$, with $G = E/[2(1+\nu)]$ being the shear modulus.

Further, unit quaternions $[q_0(s), q_1(s), q_2(s), q_3(s)]$, which satisfy $q_0^2 + q_1^2 + q_2^2 + q_3^2 = 1$, are introduced to describe the rotation of the local material frame with respect to the global material frame (Healey and Mehta, 2005; Yu and Hanna, 2019), as

$$\begin{bmatrix}\mathbf{e}_1\\ \mathbf{e}_2\\ \mathbf{e}_3\end{bmatrix} = 2\begin{bmatrix} q_0^2 + q_1^2 - \frac{1}{2} & q_1q_2 + q_0q_3 & q_1q_3 - q_0q_2\\ q_1q_2 - q_0q_3 & q_0^2 + q_2^2 - \frac{1}{2} & q_2q_3 + q_0q_1\\ q_1q_3 + q_0q_2 & q_2q_3 - q_0q_1 & q_0^2 + q_3^2 - \frac{1}{2}\end{bmatrix}\begin{bmatrix}\mathbf{E}_1\\ \mathbf{E}_2\\ \mathbf{E}_3\end{bmatrix}. \tag{4}$$

Using Eq. (4) and considering that $\mathbf{p}' = \mathbf{e}_3$, we have

$$p_1' = 2(q_1q_3 + q_0q_2),\ \ p_2' = 2(q_2q_3 - q_0q_1),\ \ p_3' = 2(q_0^2 + q_3^2) - 1. \tag{5}$$

Taking the derivative of Eq. (4) and using $\mathbf{e}_i' = \boldsymbol{\omega} \times \mathbf{e}_i$, one obtains

$$\begin{aligned} q_0' &= \left(-q_1\kappa_1 - q_2\kappa_2 - q_3\kappa_3\right)/2, \quad q_1' = \left(q_0\kappa_1 - q_3\kappa_2 + q_2\kappa_3\right)/2, \\ q_2' &= \left(q_3\kappa_1 + q_0\kappa_2 - q_1\kappa_3\right)/2, \quad q_3' = \left(-q_2\kappa_1 + q_1\kappa_2 + q_0\kappa_3\right)/2. \end{aligned} \tag{6}$$

Eqs. (3), (5), and (6) constitute thirteen ordinary differential equations (ODE) that govern the equilibrium of a Kirchhoff rod. The ODE system involves thirteen unknown variables, including the internal forces ($N_1$, $N_2$, $N_3$), curvatures ($\kappa_1$, $\kappa_2$, $\kappa_3$), position coordinates ($p_1$, $p_2$, $p_3$), and quaternions ($q_0$, $q_1$, $q_2$, $q_3$). When equipped with appropriate boundary conditions, a well-posed boundary value problem (BVP) can be defined.

As illustrated in **Fig. 1**, the rod segments are connected by rigid joints. In the rod model, these joints are represented by small, rounded corners that smoothly connect adjacent rod segments. As will be demonstrated, when the corner radius is sufficiently small, it has negligible influence on the overall mechanical behavior of the ring. However, the initial curvatures at the interfaces between the straight segments and the rounded corners are discontinuous. To address this issue, the ring is divided into multiple segments at these interfaces, with each segment modeled as a Kirchhoff rod, resulting in a multi-segment rod system. The multiple segments are coupled by imposing continuity or compatibility conditions (see **Appendix B**), producing a well-posed BVP. The BVP for the multi-segment rod system is solved using a numerical continuation method with the help of the Continuation Core and Toolboxes (COCO) (Ahsan et al., 2022; Dankowicz and Schilder, 2013) operated in MATLAB. In particular, to determine the equilibrium states of 2D segmented rings with different out-of-plane natural curvatures, the natural curvature is specified as the continuation parameter and varies from 0 to the prescribed value. To study the transition behavior of 2D segmented rings with a given natural curvature under external bending loads, the bending angle is specified as the continuation parameter and varies from 0 to $2\pi$. Details on the numerical implementation are provided in **Appendix B,** and the corresponding MATLAB code is available on GitHub[1]. By solving the multi-segment rod system, numerical solutions to the thirteen variables for each segment, corresponding to a prescribed natural curvature or bending angle, can be obtained, from which the equilibrium configurations are depicted, and the total strain energy is computed.

[1] GitHub link: https://github.com/Stanford-ZhaoLab/3D-Ring-Origami

## 3. Results and discussion

Next, the rod model combined with FEA and experiments is employed to investigate the equilibrium states and transition behavior of 2D segmented rings with out-of-plane natural curvature, using square and hexagonal rings as representative examples. Details on FEA simulations are presented in **Appendix C**. Two general cases of natural curvature distribution in the rod segments of square and hexagonal rings are discussed in detail. In the first case, all rod segments have the same positive out-of-plane natural curvature (in the $\mathbf{e}_1$-$\mathbf{e}_3$ plane), which is perpendicular to the planar ring (in the $\mathbf{e}_2$-$\mathbf{e}_3$ plane), as shown in **Figs. 2(a)** and **4(a)**, referred to as *type-I square ring* and *type-I hexagonal ring*, respectively. Note that a positive out-of-plane natural curvature induces a moment that bends the rod segment upward (toward the $\mathbf{e}_1$-direction), while a negative out-of-plane natural curvature induces a moment that bends the rod segment downward (opposite to the $\mathbf{e}_1$-direction). In the second case, the rod segments have alternating positive and negative out-of-plane natural curvatures of the same magnitude (in the $\mathbf{e}_1$-$\mathbf{e}_3$ plane), also perpendicular to the ring's planar configuration (in the $\mathbf{e}_2$-$\mathbf{e}_3$ plane), as illustrated in **Figs. 3(a)** and **5(a)**, referred to as *type-II square ring* and *type-II hexagonal ring*, respectively. The material properties of the rings are set as Young's modulus $E$ = 200 GPa and Poisson's ratio $v$ = 0.3.

In the rod model, to minimize the influence of rounded corners, the corner radius $r$ is assumed to be much smaller than the edge length $L$ of the ring. Based on a convergence study of the equilibrium paths of square and hexagonal rings with various length-to-radius ratios (see **Fig. S1** in the **Supplementary Material**), the ratio $L/r$ is taken as 100 for square rings and 50 for hexagonal rings. Moreover, the effect of corner rigidity is examined in **Figs. S2** and **S3** in the **Supplementary Material** using FEA for type-I square and hexagonal rings, respectively. It is shown that when the corners are very soft compared to the edges (e.g., $E_c = 0.01E_e$, where $E_c$ and $E_e$ represent the moduli of the corners and the edges, respectively), the ring can exhibit several highly symmetric configurations that differ from experimental observations due to weakened joint constraints. In this case, the corners undergo large deformation, including stretching, which cannot be captured by the current rod model because it neglects stretching. By contrast, for relatively stiff corners (e.g., $E_c = E_e$ and $5E_e$) and rigid corners, the ring consistently exhibits equilibrium states with reduced symmetry as we observed in experiments. In these cases, corner rigidity has little effects on the equilibrium paths and equilibrium configurations. For simplicity, we assume that the corner and edges share the same material properties in the rod model and FEA. Further, a comparative study of the equilibrium paths for different cross-sectional aspect ratios $h/t$ (see **Fig. S4** in the **Supplementary Material**) shows that multiple 3D stable states of 2D segmented rings are only possible when $h/t < 1$, meaning that the out-of-plane bending stiffness $EI_2$ must be smaller than the in-plane bending stiffness $EI_1$ of the rod segments. In this work, $h/t$ is fixed at 1/4, consistent with the value used in our experiments.

*3.1. Square rings with out-of-plane natural curvature*

As illustrated in **Fig. 1(d)**, with a given out-of-plane natural curvature, the 2D initial state of a square ring is unstable and will spontaneously snap to a 3D equilibrium state when external constraints are removed. The 3D equilibrium states of square rings with different out-of-plane natural curvatures are predicted by both the rod model and FEA simulations. **Fig. 2(b)** presents the dimensionless strain energy $UL/GJ$ as a function of the dimensionless natural curvature $\kappa_n L/2\pi$ for the equilibrium states of type-I square rings predicted by the rod model. Interestingly, the equilibrium path bifurcates into two branches when $\kappa_n L/2\pi$ increases to 0.22. In equilibrium path 1, which features symmetric deformation, all four edges of the square ring undergo identical deformation, leading to a class of highly symmetric configurations with high strain energy. In equilibrium path 2, characterized by asymmetric deformation, the symmetry of the ring is broken during deformation (only opposite edges exhibit identical deformation), resulting in a series of less symmetric configurations with significantly lower strain energy, especially when the dimensionless natural curvature is between 0.5 and 1.5. A comparison of the equilibrium configurations with symmetric and asymmetric deformations for different dimensionless natural curvatures within that range is provided in **Fig. S5** in the **Supplemental Material**. It is seen that, for the same natural curvature, equilibrium configurations with asymmetric deformations exhibit a more uniform curvature distribution along the edges, as fewer geometric symmetries constrain the system. For example, when $\kappa_n L/2\pi = 1$, the equilibrium configuration with asymmetric deformation has a uniform edge curvature equal to the natural curvature, leading to a zero-energy configuration. In contrast, the equilibrium configurations with symmetric deformation have nonuniform curvature in each edge, resulting in configurations with high energy. As a result, preserving symmetry requires more energy, making asymmetric deformation energetically favorable. In other words, type-I square rings prefer to equilibrate in a configuration with asymmetric deformation. This symmetry breaking mechanism also appears in rings with other geometries, including triangular and pentagonal rings (**Figs. S6** and **S7** in the **Supplementary Material**) and hexagonal rings (**Fig. 4**), highlighting the importance of geometric symmetry in shaping energy landscapes and governing equilibrium configurations of ring systems with out-of-plane natural curvature. Note that in the rod model, equilibrium path 2 is triggered by introducing a small geometric imperfection to break symmetry. Specifically, the length $l_1$ of one pair of edges is set to be 0.1% longer than the length $l_2$ of the other pair (i.e., $l_1 = 1.001l_2$).

The equilibrium path 2 predicted by the rod model is compared with the FEA result in **Fig. 2(c)**. A good agreement is observed between the two methods, except in the two shaded regions, where the rod model predicts an equilibrium path with snap-back feature, while the FEA result shows a sudden energy drop. This difference is attributed to the distinct methods used to obtain the equilibrium solutions: a numerical continuation method is employed in the rod model, and a temperature gradient-controlled loading method is used in FEA simulation (see **Appendix C** for details). The first shaded region corresponds to a dimensionless natural curvature range of (0.46, 0.52), within which the rod model predicts three equilibrium states for a given natural curvature. For example, when $\kappa_n L/2\pi = 0.5$, configurations of the three equilibrium states corresponding to the dots in the enlarged inset are shown in **Fig. 2(d)**. In particular, equilibrium states ① and ③ have the same strain energy. However, state ① lies on the rising branch of the equilibrium path, while state ③ lies on the descending branch. As demonstrated in **Figs. 2(e)&(f)** and **Fig. S8** in the **Supplementary Material**, equilibrium states ① and ③ are stable, whereas equilibrium state ② is unstable, as it corresponds to a local energy maximum. Additionally, under external bending loads, two more stable states having the same configurations as equilibrium states ① and ③ are triggered, thereby enabling multistability. The second shaded region is extremely narrow, corresponding to a dimensionless natural curvature range of (1.49, 1.50), and also enables multistability, as demonstrated in **Fig. S9** in the **Supplementary Material**. Moreover, when the dimensionless natural curvature is equal to a positive integer (e.g., 1 and 2), the equilibrium state exhibits a spherical configuration formed by two orthogonal circles (**Fig. 2(d)**). These circles are multilayered (2-layer overlapping for $\kappa_n L/2\pi = 1$ and 4-layer overlapping for $\kappa_n L/2\pi = 2$) and each has an edge curvature equal to the natural curvature. As a result, these equilibrium configurations have zero strain energy. Experimental demonstrations of the various equilibrium configurations are also shown in **Fig. 2(d)**, except for the unstable equilibrium configuration ② at $\kappa_n L/2\pi = 0.5$.

Good agreements are observed between the experimental configurations and the theoretical predictions.

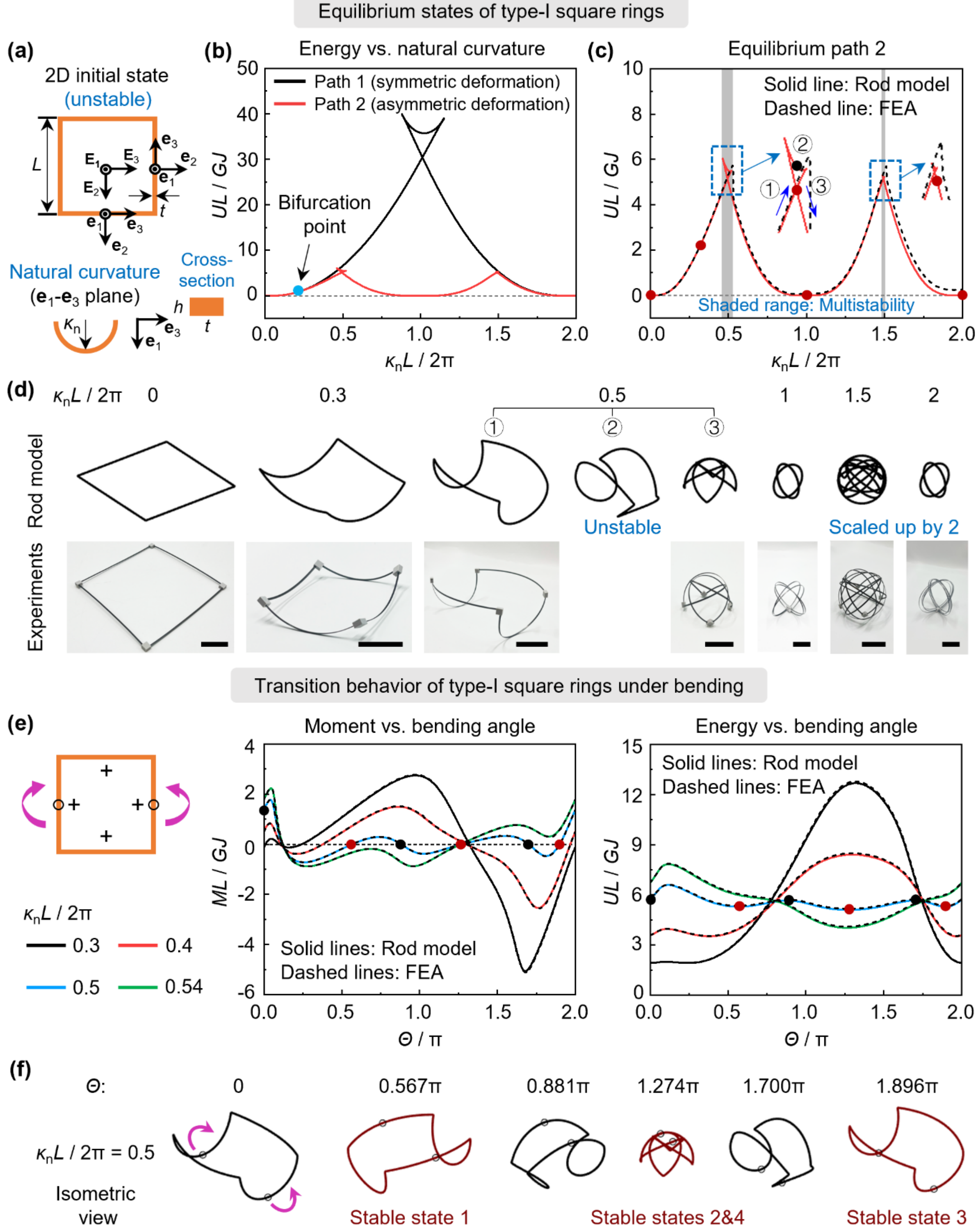

**Fig. 2.** Equilibrium states and transition behavior of type-I square rings. (a) Schematic of the 2D initial state. The global material frame ($\mathbf{E}_1$, $\mathbf{E}_2$, $\mathbf{E}_3$) is located at the center of the ring, while the local material frame ($\mathbf{e}_1$, $\mathbf{e}_2$, $\mathbf{e}_3$) is attached to the centerline of each rod segment. The natural curvature lies in the $\mathbf{e}_1$-$\mathbf{e}_3$ plane, which is perpendicular to the $\mathbf{e}_2$-$\mathbf{e}_3$ plane of the initial ring configuration. The edges marked in orange have the same positive natural curvature, which induces a moment that bends the rod segment upward, toward the $\mathbf{e}_1$-direction. (b) Dimensionless strain energy $UL/GJ$ versus dimensionless natural curvature $\kappa_n L/2\pi$ for two distinct equilibrium paths predicted by the rod model. (c) Comparison of the equilibrium path 2 predicted by the rod model and FEA. The dots represent the locations of the equilibrium configurations presented in panel (d), with dark red dots indicating stable states and the black dot indicating an unstable state. The shaded ranges denote regions of natural curvature where the ring exhibits multistability. (d) Representative equilibrium configurations for various dimensionless natural curvatures predicted by the rod model and validated by experiments. Scale bars: 4 cm. (e) Dimensionless moments $ML/GJ$ and strain energy $UL/GJ$ as functions of the normalized bending angle $\Theta/\pi$ for type-I square rings with different dimensionless natural curvatures $\kappa_n L/2\pi$ under external bending loads. Circles on the ring schematic denote the loading positions. Dark red dots on the moment and energy curves represent local energy minima, while black dots (except for the first one) indicate local energy maxima. (f) Selected configurations corresponding to the dots on the moment and energy curves in panel (e) during the transition of a type-I square ring with $\kappa_n L/2\pi = 0.5$ under increasing bending loads.

The transition behavior of type-I square rings with different natural curvatures within or near the first shaded range (0.46, 0.52), subjected to a pair of bending loads applied at the midpoints of opposite edges, is examined in **Fig. 2(e)** using both the rod model and FEA. It can be observed that the variations in dimensionless moment $ML/GJ$ and strain energy $UL/GJ$ as functions of the normalized bending angle $\Theta/\pi$ predicted by the two methods agree well for various natural curvatures. When the natural curvature falls within the range corresponding to the first shaded region (e.g., $\kappa_n L/2\pi = 0.5$), there are three local minima (represented by dark red dots) on the energy landscape of the square ring during the transition process, indicating the presence of three stable states. The transition process among these three stable states under inward bending is presented in **Fig. 2(f)**. As the bending angle increases, the square ring first transitions to stable state 1 with a tent-like configuration, then to stable state 2 featuring a dome-like shape, and eventually to stable state 3, which also exhibits a tent-like configuration. Note that the transition from stable state 2 to stable state 3 requires edge penetration (in the rod model and FEA, edge contact and penetration are not considered), which is infeasible in experiments. Alternatively, the transition between the two stable states can be achieved by simply reversing the bending direction. As shown in **Fig. S8** in the **Supplementary Material**, when bending outward, the square ring first transitions to the tent-like stable state 3 and then to the dome-like stable state 4. As illustrated in **Fig. 1(d)**, stable states 1 and 3 (as well as stable states 2 and 4) share the same configuration, while the same rod segments in the two states undergo different deformations (or stack in different ways). Therefore, they are regarded as four different stable states. An experimental demonstration of the transition between the four stable states is provided in **Video 2** in the **Supplementary Material**. When the natural curvature is below the shaded range (e.g., $\kappa_n L/2\pi$ = 0.3 and 0.4), the square ring exhibits two stable states with identical configurations, where the two pairs of edges are curved to different extents similar to stable state 1 for $\kappa_n L/2\pi = 0.5$. When the natural curvature slightly exceeds the shaded range (e.g., $\kappa_n L/2\pi = 0.54$), the square ring becomes monostable and remains stable only in the dome-like configuration. For type-I square rings with dimensionless natural curvatures greater than 1, **Fig. S9** in the **Supplementary Material** shows FEA results of their transition behavior under bending loads. These results indicate that multistability only occurs when the natural curvature is within the shaded region. In all other cases, the square ring is monostable.

Equilibrium states and transition behavior of type-II square rings are examined in **Fig. 3**, where the rod segments possess alternating positive and negative out-of-plane natural curvatures of equal magnitude (**Fig. 3(a)**). In this case, the rod model also predicts two equilibrium paths corresponding to symmetric and asymmetric deformations, respectively, as the natural curvature increases (the equilibrium path associated with asymmetric deformation is also triggered by introducing a small geometric imperfection), as shown in **Fig. 3(b)**. Comparisons of the equilibrium configurations between these two paths are presented in **Fig. S10** in the **Supplementary Material**. Similar to the case of type-I square rings, the equilibrium path associated with asymmetric deformation possesses significantly lower strain energy than the symmetric counterpart. Along this path, type-II square rings feature three distinct equilibrium states for a given natural curvature within two specific ranges: a broader range of (0.47, 0.53), and a narrower range of (1.51, 1.53), as illustrated by the shaded regions in **Fig. 3(c)**. Also, type-II square rings with natural curvature within these shaded ranges exhibit multistability, as demonstrated in **Figs. 3(e)&(f)** and **Fig. S11** in the **Supplementary Material**. The three equilibrium configurations corresponding to $|\kappa_n|L/2\pi = 0.5$ are shown in **Fig. 3(d)**, along with additional equilibrium configurations at other natural curvatures. Both theoretical predictions and experimental results are presented. Due to the opposite directions of the natural curvature-induced bending moments in the two pairs of edges, type-II square rings give rise to a new class of 3D equilibrium configurations. For example, when $|\kappa_n|L/2\pi = 1$ or 2, type-II square ring exhibits a 3D figure-eight configuration consisting of two tangent circles with zero strain energy. As validated by experiments, all these equilibrium configurations are stable, except for the equilibrium configuration ② at $|\kappa_n|L/2\pi = 0.5$, which corresponds to a local energy maximum state (see **Figs. 3(e)&(f)**).

**Fig. 3(e)** presents the moment and energy evolution during the state transitions of type-II square rings with several natural curvatures within or near the first shaded range of (0.47, 0.53) subjected to bending loads. When the dimensionless natural curvature is within this range (e.g., $|\kappa_n|L/2\pi = 0.5$), the square ring exhibits three stable states, indicated by dark red dots on the moment and energy curves. The corresponding transition process is illustrated in **Fig. 3(f)** and experimentally demonstrated in **Video 3** in the **Supplementary Material**. It is seen that equilibrium configurations ① and ③ shown in **Fig. 3(d)** correspond to stable states 1 and 2, respectively, while equilibrium configuration ② is unstable and associated with a local energy maximum that separates the two stable states. Upon the application of external bending, a third stable state (i.e., stable state 3), identical in configuration to stable state 1, is triggered. Notably, transitions among the three stable states occur without edge contact or penetration, and reversing the bending direction leads to the same three stable states (see **Fig. S12** and **Video 3** in the **Supplementary Material**). When the dimensionless natural curvature lies outside the shaded range, such as $|\kappa_n|L/2\pi = 0.3$, 0.4, and 0.54 studied in **Fig. 3(e)**, or larger values considered in **Fig. S11** in the **Supplementary Material**, the square ring exhibits either monostability or bistability. An additional example of square rings with out-of-plane natural curvature is presented in **Fig. S13** in the **Supplementary Material**, where two adjacent rod segments of the square ring have positive natural curvature while the other two have negative natural curvature (referred to as *type-III square ring*). The type-III square ring shows another new class of equilibrium states as the natural curvature increases. Similar to type-I and type-II square rings, it exhibits multistability within two specific ranges of natural curvature, while displaying monostability or bistability outside these ranges.

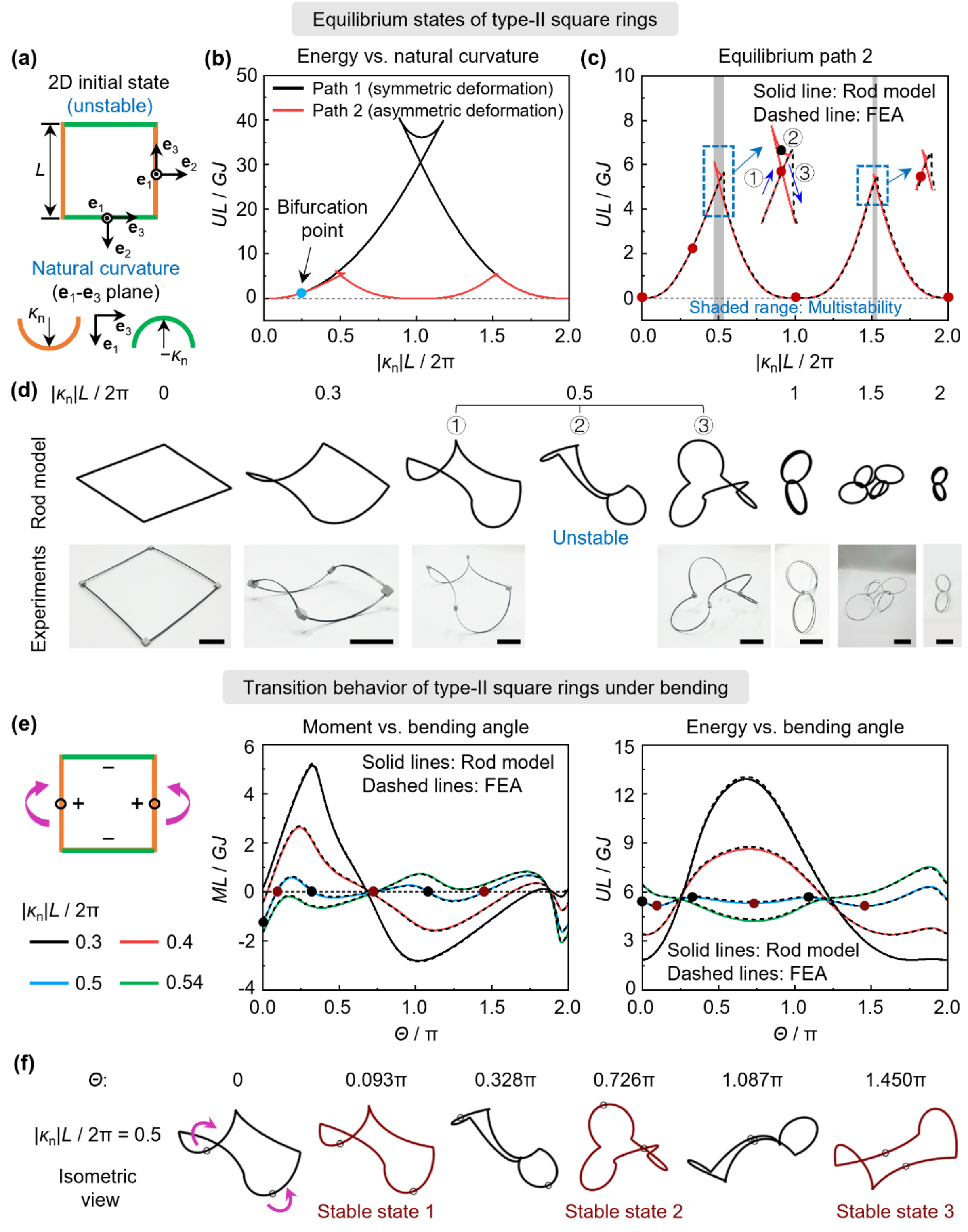

**Fig. 3.** Equilibrium states and transition behavior of type-II square rings. (a) Schematic of the initial 2D state. The natural curvature lies in the $\mathbf{e}_1$-$\mathbf{e}_3$ plane, which is perpendicular to the $\mathbf{e}_2$-$\mathbf{e}_3$ plane of the initial ring configuration. The edges marked in orange have the same positive natural curvature, which induces a moment that bends the rod segment upward, toward the $\mathbf{e}_1$-direction. The edges marked in green have the same negative natural curvature, which induces a moment that bends the rod segment downward, opposite to the $\mathbf{e}_1$-direction. (b) Dimensionless strain energy $UL/GJ$ versus dimensionless natural curvature $|\kappa_n|L/2\pi$ for two distinct equilibrium paths predicted by the rod model. (c) Comparison of the equilibrium path 2 predicted by the rod model and FEA. The dots represent the locations of the equilibrium configurations presented in panel (d), with dark red dots indicating stable states and the black dot indicating an unstable state. The shaded ranges denote regions of natural curvature where the ring exhibits multistability. (d) Representative equilibrium configurations for various dimensionless natural curvatures predicted by the rod model and validated by experiments. Scale bars: 4 cm. (e) Dimensionless moments $ML/GJ$ and strain energy $UL/GJ$ as functions of the normalized bending angle $\Theta/\pi$ for type-II square rings with different dimensionless natural curvatures $|\kappa_n|L/2\pi$ under external bending loads. Circles on the ring schematic denote the loading positions. Dark red dots on the moment and energy curves represent local energy minima, while black dots (except for the first one) indicate local energy maxima. (f) Selected configurations corresponding to the dots on the moment and energy curves in panel (e) during the transition of a type-II square ring with $|\kappa_n|L/2\pi = 0.5$ under increasing bending loads.

### *3.2. Hexagonal rings with out-of-plane natural curvature*

In this subsection, we discuss the equilibrium states and transition behavior of hexagonal rings with out-of-plane natural curvature. As introduced at the beginning of **Section 3**, two types of hexagonal rings are discussed in detail: *type-I hexagonal rings*, composed of rod segments with the same positive out-of-plane natural curvature (**Fig. 4(a)**), and *type-II hexagonal rings*, consisting of rod segments with alternating positive and negative out-of-plane natural curvatures of equal magnitude (**Fig. 5(a)**). The two equilibrium paths for type-I hexagonal rings predicted by the rod model are shown in **Fig. 4(b)**. Similar to the case of type-I square rings, type-I hexagonal rings have two distinct equilibrium paths as the natural curvature increases. In equilibrium path 1, all edges undergo the same deformation, resulting in a centrally symmetric configuration with high strain energy. In contrast, in equilibrium path 2, the three pairs of edges experience different deformations, leading to a configuration with reduced symmetry but significantly lower strain energy. Comparisons of the equilibrium configurations along the two paths for various dimensionless natural curvatures are provided in **Fig. S14** in the **Supplementary Material**. Equilibrium path 2, characterized by asymmetric deformation, is further validated by FEA in **Fig. 4(c)**, and a good agreement is observed between the two methods except in the shaded ranges, where the rod model predicts three distinct equilibrium states for a given natural curvature. As explained previously, this discrepancy is due to the different methods used to determine the equilibrium solutions. For type-I hexagonal rings, four such ranges are identified as the dimensionless natural curvature increases from 0 to 2, and these ranges are considerably wider than those observed in square rings. From low to high, the dimensionless natural curvatures for the four shaded ranges are (0.19, 0.36), (0.58, 0.83), (1.18, 1.38), and (1.59, 1.82). As demonstrated in **Figs. 4(e)&(f)** and **Fig. S15** in the **Supplementary Material**, type-I hexagonal rings with natural curvature within these shaded ranges exhibit multistability. Several representative equilibrium configurations of type-I hexagonal rings are illustrated in **Fig. 4(d)**, with their corresponding locations on equilibrium path 2 indicated by dots in **Fig. 4(c)**. In particular, when the dimensionless natural curvature is an integer multiple of 0.5, i.e., $\kappa_n L/2\pi = 0.5n$ ($n$ = 1, 2, 3, ⋯), the type-I hexagonal ring exhibits 3D stable configurations with zero strain energy. Among these, the first zero-strain energy configuration ($n$ = 1) has a 3D heart-like shape, while the remaining ones ($n \geq 2$) exhibit spherical shapes. Experimental images of these equilibrium configurations are also shown in **Fig. 4(d)**. Experimental results indicate that all these equilibrium configurations are stable, except for the two configurations labeled ②, corresponding to $\kappa_n L/2\pi$ = 0.29 and 0.7 within the first and second shaded ranges, respectively. Similar to the case of square rings, these two configurations are in states of local energy maxima, as evidenced by the transition behavior under bending loads, as shown in **Figs. 4(e)&(f)** and **Fig. S15** in the **Supplementary Material**.

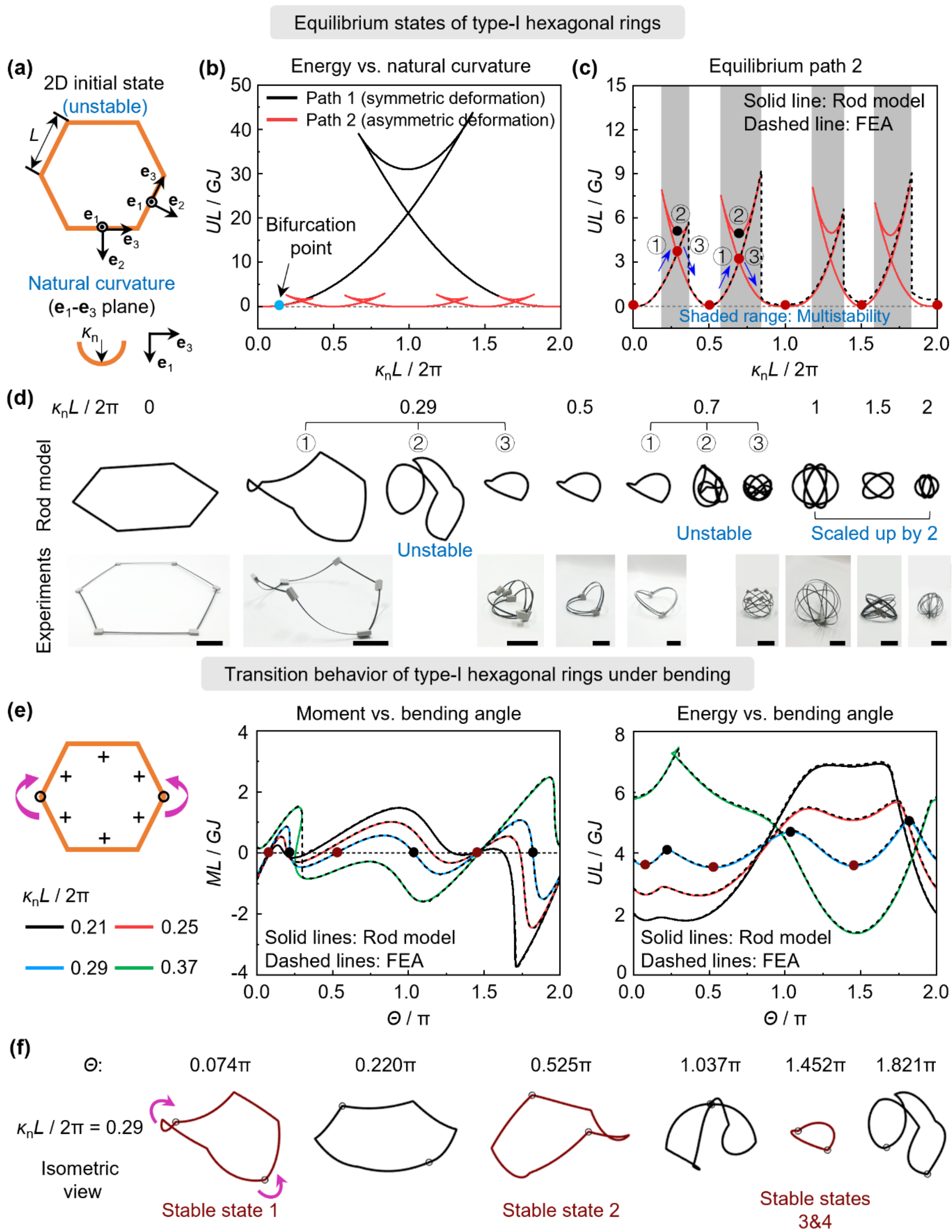

**Fig. 4.** Equilibrium states and transition behavior of type-I hexagonal rings. (a) Schematic of the initial 2D state. The natural curvature lies in the $\mathbf{e}_1$-$\mathbf{e}_3$ plane, which is perpendicular to the $\mathbf{e}_2$-$\mathbf{e}_3$ plane of the initial ring configuration. The edges marked in orange have the same positive natural curvature, which induces a moment that bends the rod segment upward, toward the $\mathbf{e}_1$-direction. (b) Dimensionless strain energy $UL/GJ$

versus dimensionless natural curvature $\kappa_n L/2\pi$ for two distinct equilibrium paths predicted by the rod model. (c) Comparison of the equilibrium path 2 predicted by the rod model and FEA. The dots represent the locations of the equilibrium configurations presented in panel (d), with dark red dots indicating stable states and black dots indicating unstable states. The shaded ranges denote regions of natural curvature where the ring exhibits multistability. (d) Representative equilibrium configurations for various dimensionless natural curvatures predicted by the rod model and validated by experiments. Scale bars: 4 cm. (e) Dimensionless moments $ML/GJ$ and strain energy $UL/GJ$ as functions of the normalized bending angle $\Theta/\pi$ for type-I hexagonal rings with different dimensionless natural curvatures $\kappa_n L/2\pi$ under external bending loads. Circles on the ring schematic denote the loading positions. Dark red dots on the moment and energy curves represent local energy minima, while black dots indicate local energy maxima. (f) Selected configurations corresponding to the dots on the moment and energy curves in panel (e) during the transition of a type-I hexagonal ring with $\kappa_n L/2\pi = 0.29$ under increasing bending loads.

The transition behavior of type-I hexagonal rings with various out-of-plane natural curvatures within or near the first shaded range of (0.19, 0.36), subjected to a pair of bending loads at corners, is studied in **Fig. 4(e)**. It can be observed that, when $\kappa_n L/2\pi = 0.29$, the energy landscape during the state transition reveals three local minima (represented by dark red dots), indicating the presence of three stable states. As expected, equilibrium configurations ① and ③ correspond to stable states 1 and 3, respectively, while equilibrium configuration ② is unstable and corresponds to a local energy maximum state at a bending angle $\Theta = 1.821\pi$ (**Fig. 4(f)**). Interestingly, a new stable configuration (i.e., stable state 2) is triggered during the transition process. It should be stated that by reversing the bending direction, stable state 1 directly transitions into stable state 4, which shares the same configuration as stable state 3 but the same rod segments in the two states stack in different ways (see **Fig. S16** and **Video 4** in the **Supplementary Material**). As a result, the type-I hexagonal ring with $\kappa_n L/2\pi = 0.29$ exhibits four stable states, and an experimental demonstration of the transition among these four states is shown in **Video 4** in the **Supplementary Material**. Multistability is also observed for other values of dimensionless natural curvature within the shaded ranges, such as $\kappa_n L/2\pi = 0.21$ and 0.25 (within the first shaded range, see **Fig. 4(e)**), and 0.7, 1.3, and 1.7 (within the second, third, and fourth shaded ranges, respectively, see **Fig. S15** in the **Supplementary Material**). These observations confirm that the type-I hexagonal ring exhibits multistability when the dimensionless natural curvature falls within the shaded ranges in **Fig. 4(c)**. In contrast, when the dimensionless natural curvature is outside these ranges, the ring becomes either monostable or bistable, as demonstrated in **Fig. 4(e)** for $\kappa_n L/2\pi = 0.37$ and **Fig. S17** in the **Supplementary Material** for $\kappa_n L/2\pi = 0.5$, 1, 1.5, and 2.

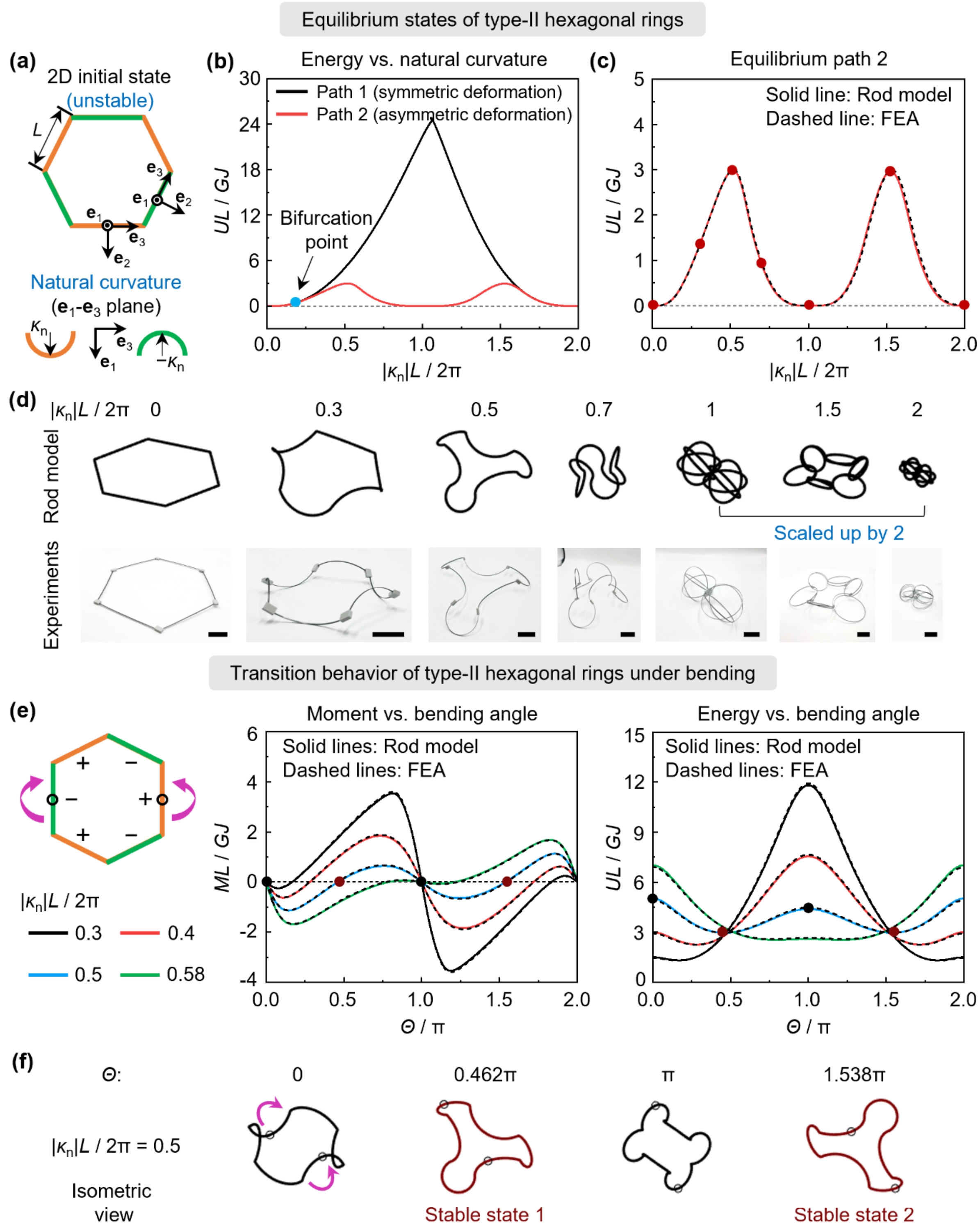

**Fig. 5.** Equilibrium states and transition behavior of type-II hexagonal rings. (a) Schematic of the initial 2D state. The natural curvature lies in the $\mathbf{e}_1$-$\mathbf{e}_3$ plane, which is perpendicular to the $\mathbf{e}_2$-$\mathbf{e}_3$ plane of the initial ring configuration. The edges marked in orange have the same positive natural curvature, which induces a moment that bends the rod segment upward, toward the $\mathbf{e}_1$-direction. The edges marked in green have the same negative natural curvature, which induces a moment that bends the rod segment downward, opposite

to the $\mathbf{e}_1$-direction. (b) Dimensionless strain energy $UL/GJ$ versus dimensionless natural curvature $|\kappa_n|L/2\pi$ for two distinct equilibrium paths predicted by the rod model. (c) Comparison of the equilibrium path 2 predicted by the rod model and FEA. Dark red dots represent the locations of the equilibrium configurations presented in panel (d). (d) Representative equilibrium configurations for various dimensionless natural curvatures predicted by the rod model and validated by experiments. Scale bars: 4 cm. (e) Dimensionless moments $ML/GJ$ and strain energy $UL/GJ$ as functions of the normalized bending angle $\Theta/\pi$ for type-II hexagonal rings with different dimensionless natural curvatures $|\kappa_n|L/2\pi$ under external bending loads. Circles on the ring schematic denote the loading positions. Dark red dots on the moment and energy curves represent local energy minima, while black dots indicate local energy maxima. (f) Selected configurations corresponding to the dots on the moment and energy curves in panel (e) during the transition of a type-II hexagonal ring with $|\kappa_n|L/2\pi = 0.5$ under increasing bending loads.

Equilibrium states and transition behavior of type-II hexagonal rings are studied in **Fig. 5**. Type-II hexagonal rings also have two equilibrium paths: path 1 featuring symmetric deformation, leading to high strain energy, and path 2 characterized by asymmetric deformation, resulting in significantly lower strain energy (see **Fig. 5(b)**). Comparisons of the equilibrium configurations associated with these two paths are provided in **Fig. S18** in the **Supplementary Material**. As shown in **Fig. 5(c)**, unlike the cases discussed in **Figs. 2-4**, the equilibrium path 2 of type-II hexagonal rings is smooth, with no ranges where three equilibrium states coexist for a given natural curvature. This indicates that type-II hexagonal rings do not exhibit multistability. The equilibrium configurations corresponding to the dark red dots on equilibrium path 2 in **Fig. 5(c)** are illustrated in **Fig. 5(d)**, as predicted by the rod model and validated by experiments. All these configurations are stable. Similar to type-II square rings, type-II hexagonal rings form 3D figure-eight configurations with zero strain energy when the dimensionless natural curvature is a positive integer (e.g., $|\kappa_n|L/2\pi = 1$ and 2). Transition behavior of type-II hexagonal rings with various natural curvatures under bending loads is further examined in **Fig. 5(e)** and **Fig. S19** in the **Supplementary Material**. These results reveal that type-II hexagonal rings can only be monostable or bistable. In the bistable cases, the two stable states share the same configuration and transition between each other through inversion, as demonstrated for $|\kappa_n|L/2\pi = 0.5$ in **Fig. 5(f)**, with an experimental demonstration provided in **Video 5** in the **Supplementary Material**. **Fig. S20** in the **Supplementary Material** provides another example of hexagonal rings with out-of-plane natural curvature, referred to as *type-III hexagonal rings*. The type-III hexagonal rings have one pair of opposite edges with positive natural curvature and the other two pairs of opposite edges with negative natural curvature, which achieve a series of equilibrium configurations different from those of type-I and type-II hexagonal rings as the natural curvature increases. Similar to type-II hexagonal rings, type-III hexagonal rings are found to be either monostable or bistable. This phenomenon differs from that of square rings, where both type-II and type-III square rings can still exhibit multistability within certain ranges of natural curvature. The difference arises from their distinct geometries. In a square ring, there are only four joints with right angles, and the relatively simple geometry with fewer constraints allows it not to create excessive geometric incompatibility when negative natural curvatures are introduced (e.g., type-II and type-III square rings), thereby preserving multistability. In contrast, a hexagonal ring has six joints with 120° angles, and the relatively complex geometry and more constraints amplifies the geometric incompatibility introduced by negative natural curvatures (e.g., type-II and type-III hexagonal rings). As a result, the multistability present in type-I hexagonal rings is lost in the type-II and type-III cases. This geometric sensitivity also suggests that polygonal rings with all edges having the same out-of-plane natural curvature tend to exhibit multistability (within certain natural curvature range) across different topologies, whereas rings with both positive and negative out-of-plane natural curvatures may preserve multistability only in relatively simple geometries (e.g., squares).

In **Figs. 1-5**, the natural curvature of the rod segments lies in a plane that is perpendicular to the ring's planar configuration, inducing a moment that bends the rod segment either upward or downward. In practice, the direction of the natural curvature-induced bending moment can be altered to other directions by adjusting the edge orientations. Experimental demonstrations to achieve this are shown in **Fig. 6**, where the slot orientation at the joint used to connect adjacent rod segments is prescribed before 3D printing. The slot orientation angles in the two orthogonal planes are denoted by $\theta_1$ and $\theta_2$ (**Fig. 6(a-i)**). As illustrated in **Fig. 6(b)**, when $\theta_1 = \theta_2 = 45°$, the square ring composed of stainless-steel rod segments having the same dimensionless natural curvature $\kappa_n L/2\pi = 0.5$ (**Fig. 6(a-ii)**) exhibits a single stable state. However, when $\theta_1 = 90°$ and $\theta_2 = 45°$, the ring formed by the same rod segments becomes bistable (**Fig. 6(c)**). This is in sharp contrast to the case presented in **Fig. 1(d)**, where the square ring with $\kappa_n L/2\pi = 0.5$ and $\theta_1 = \theta_2 = 90°$ exhibits four stable states. Therefore, tuning the edge orientations of the rod segments offers an additional approach to program the 3D stable states of 2D segmented rings. This also indicates that the joint tolerances (e.g., small angular deviations in the slots) may affect the ring's stability. Moreover, because the multistability occurs only within relatively narrow ranges of natural curvature, particularly for square rings, rod bending accuracy during fabrication is also critical.

The influences of these imperfections can be mitigated by improving fabrication precision, for example by directly 3D printing the joints with predesigned slots and the rod segments in their natural curved states, followed by assembly.

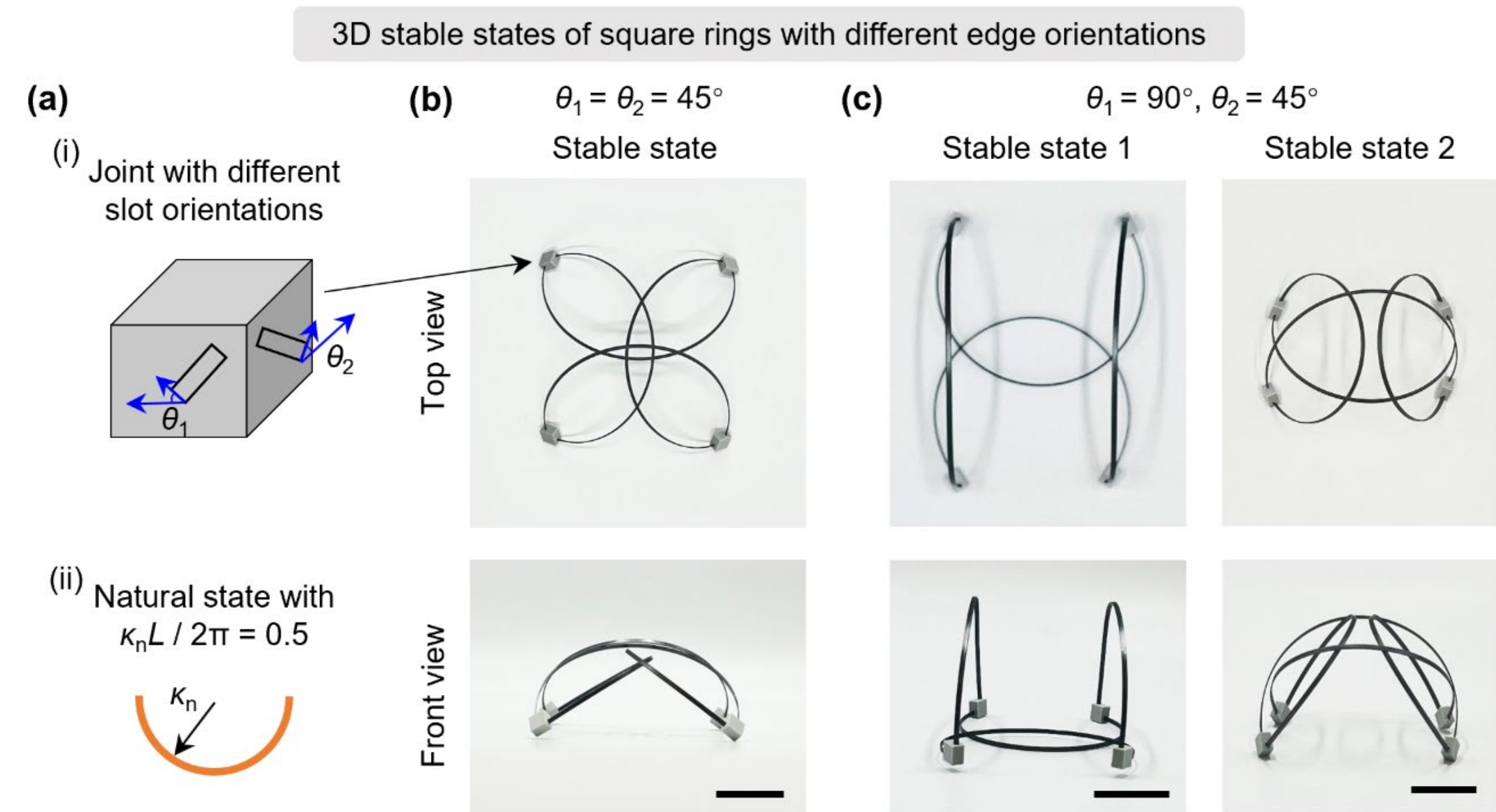

**Fig. 6.** 3D stable states of square rings with different edge orientations. (a-i) Schematic of a joint with different slot orientation angles $\theta_1$ and $\theta_2$ in two orthogonal planes, and (a-ii) schematic of the natural state of a rod segment with dimensionless natural curvature $\kappa_n L/2\pi = 0.5$. (b) Stable state of a square ring composed of four identical stainless-steel rod segments with $\kappa_n L/2\pi = 0.5$, assembled by four joints with $\theta_1 = \theta_2 = 45°$. (c) Stable states of a square ring composed of four identical stainless-steel rod segments with $\kappa_n L/2\pi = 0.5$, assembled by four joints with $\theta_1 = 90°$ and $\theta_2 = 45°$. Scale bars: 3 cm.

## 4. Concluding remarks

In summary, we have demonstrated that 2D ring origami consisting of closed-loop segmented rods can transform into diverse 3D equilibrium states by introducing out-of-plane natural curvature (i.e., the stress-free curved state lies in a plane perpendicular to the planar ring) into the rod segments. The equilibrium states and transition behavior of these segmented rings were predicted by a multi-segment Kirchhoff rod model, and its validity was confirmed through both finite element simulations and experiments. Using square and hexagonal rings as representative examples, we systematically investigated the effects of out-of-plane natural curvature on equilibrium states and transition behavior of 2D segmented rings. Our findings indicate that the magnitude of the natural curvature, the direction of the natural curvature-induced bending moment, and the number of rod segments all play significant roles in the mechanical behavior of these rings. By strategically programming these parameters, 2D segmented rings can achieve various intriguing and functional features, including geometric simplicity, programmable stability, and spontaneous 2D-to-3D transformation. For instance, segmented rings composed of rod segments having identical out-of-plane natural curvature exhibit multistability with various 3D configurations within specific natural curvature ranges, while showcasing monostability but with a compact zero-energy spherical configuration when the dimensionless natural curvature $\kappa_n L/2\pi$ (where $L$ is the length of the rod segment) is a positive integer (e.g., 1 and 2). This property enables the construction of complex shape-morphing structures from simple geometries. Moreover, the initial 2D state of these rings are intrinsically unstable, allowing spontaneous transitions into stable 3D states through snap-folding instabilities. Taken together, these merits make segmented rings particularly promising for applications such as deployable aerospace structures, where high packing ability and easy-to-actuate shape morphing are highly desired.

Additionally, the natural curvature and length of individual rod segments as well as the segment number can be independently programmed, offering a vast design space for 2D segmented rings to achieve diverse 3D equilibrium configurations and mechanical behavior (Liu et al., 2020; Zhang et al., 2022b). When integrated with optimization algorithms or machine learning techniques, they can serve as a versatile platform for constructing structures to achieve target stable configurations or mechanical responses. It should be noted that the design space of the proposed ring origami can be further broadened to other material systems. For example, the natural curvature-enabled 2D-to-3D transformation can be realized using thermally responsive materials with nonuniform thermal expansion (Bae et al., 2014; Maurin et al., 2024), inflatable structures with nonuniform inflation (Jin et al., 2020; Tanaka et al., 2023), or magnetic soft materials with rationally programmed magnetization profiles (Kim et al., 2018; Wu et al., 2020), where external stimuli (e.g., thermal, pneumatic, or magnetic fields) induce out-of-plane moments that trigger the shape transformation. In addition, transitions between different stable states can be achieved by integrating active materials, such as liquid crystal elastomers (Kim et al., 2023; Zhao et al., 2023), shape memory polymers (Wan et al., 2024; Ze et al., 2020), and magnetic soft materials (Kim et al., 2018; Leanza et al., 2024a), with appropriate structural designs, where external actuation produces bending moments that drive the state transition.

## Acknowledgments

This work was supported by the National Science Foundation Award CPS-2201344 and National Science Foundation Career Award CMMI-2145601.

## Data availability

MATLAB code for the numerical implementation of the multi-segment rod model is available on GitHub: https://github.com/Stanford-ZhaoLab/3D-Ring-Origami

## Appendix A. Experiments

The rings are composed of rod segments made of stainless steel, which are manually assembled through rigid joints. These joints, which incorporate predesigned slots for connecting the rod segments, are 3D printed using Polylactic acid (PLA). The slot orientations can be adjusted as needed to tune the direction of natural curvature-induced bending moment, thereby altering the resulting stable states (see schematic of the joint with slots in **Fig. 6(b)**). Moreover, the stainless-steel rods can be plastically deformed by hand to achieve prescribed values of the natural curvature. In all experiments, the rods have the same cross-sectional width $t$ = 2 mm and height $h$ = 0.5 mm, and the length varies as necessary for different values of natural curvatures considered. For example, in **Figs. 2-5**, shorter rods are used to fabricate the rings with smaller dimensionless natural curvatures, whereas longer rods are used for larger natural curvatures. This demonstrates that the proposed design strategy is scalable and can be scaled up or down for different application scenarios. To enhance durability in practical applications, materials with high fatigue resistance (e.g., carbon-fiber reinforced polymers) can be used to construct the rings. Additionally, alternative fabrication methods are feasible, such as directly 3D printing the rod segments in their natural curved states and then assembling them, which can also improve fabrication precision.

## Appendix B. Numerical implementation for the multi-segment rod model

In the present work, square and hexagonal rings are studied as representative examples. The rod segments of the square and hexagonal rings either have the same out-of-plane natural curvatures (referred to as type-I square and hexagonal rings) or have alternating positive and negative out-of-plane natural curvatures of the same magnitude (referred to as type-II square and hexagonal rings). As illustrated in **Fig. B1**, due to geometric symmetry, the analysis of the type-I and type-II square rings, as well as the type-I hexagonal ring, can be simplified by considering only one quarter of the structure, while the type-II hexagonal ring can be reduced to one half. The selected portions (indicated by dashed boxes) are further divided into three, three, four, and seven segments, respectively, at the interface between straight edges and rounded corners. Each segment is modeled as a Kirchhoff rod, resulting in a multi-segment rod system.

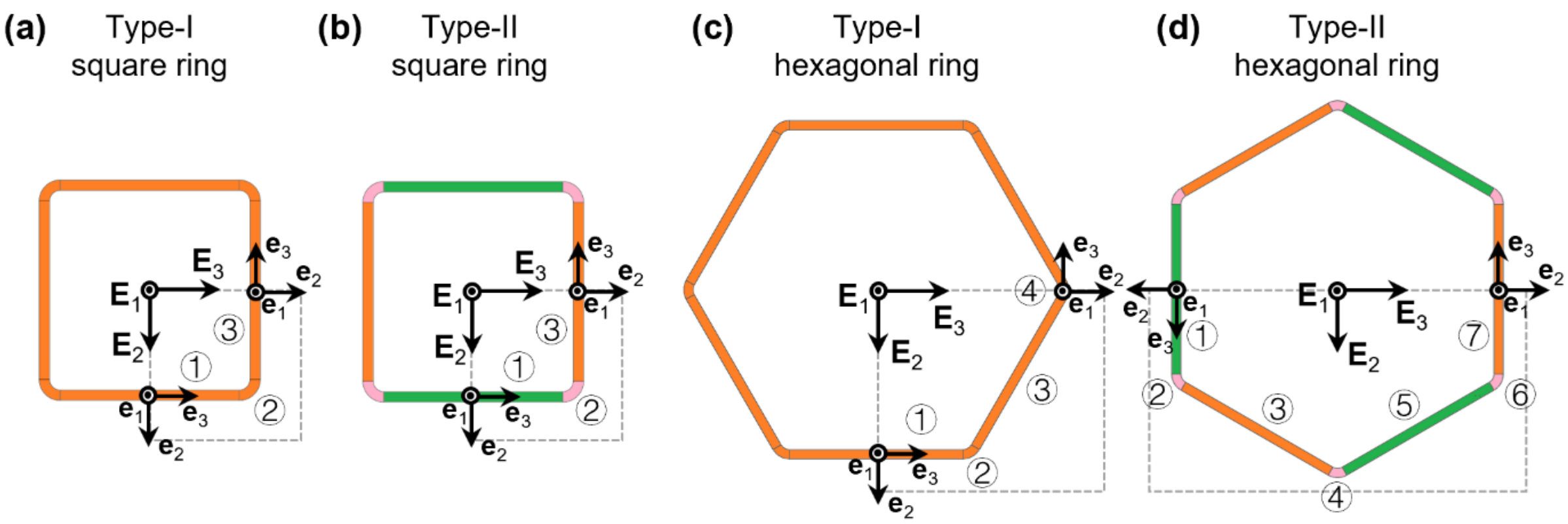

**Fig. B1.** Schematics of the square and hexagonal rings with out-of-plane natural curvature considered in the rod model. (a) Type-I square ring. (b) Type-II square ring. (c) Type-I hexagonal ring. (d) Type-II hexagonal ring. The global material frame ($\mathbf{E}_1$, $\mathbf{E}_2$, $\mathbf{E}_3$) is fixed at the center of the ring, and the local material frame ($\mathbf{e}_1$, $\mathbf{e}_2$, $\mathbf{e}_3$) is attached to the centerline of the ring's edge. The orange edges and corners have a positive out-of-plane natural curvature $\kappa_n$, the green edges have a negative out-of-plane natural curvature $-\kappa_n$, and the pink corners are assumed to have a nonuniform natural curvature which linearly varies from $\kappa_n$ to $-\kappa_n$ or from $-\kappa_n$ to $\kappa_n$. Due to symmetry, one quarter of the type-I and type-II square rings and the type-I hexagonal ring, and one half of the type-II hexagonal ring are used for the analysis. These portions (indicated by dashed boxes) are further divided into multiple segments, and each segment is labeled by a circled number.

### *B.1. Dimensionless governing equations*

The governing equations of the multi-segment rod system is first nondimensionalized by introducing the following dimensionless quantities,

$$\overline{N}_i^{(j)} = \frac{N_i^{(j)} r^2}{GJ},\ (\overline{\kappa}_i^{(j)}, \overline{\kappa}_0^{(j)}, \overline{\kappa}_\mathrm{n}^{(j)}) = (\kappa_i^{(j)}, \kappa_0^{(j)}, \kappa_\mathrm{n}^{(j)}) r,\ \overline{p}_i^{(j)} = \frac{p_i^{(j)}}{r},$$
$$\overline{s}^{(j)} = \frac{s^{(j)}}{r\xi^{(j)}},\ \frac{d(\bullet)}{ds^{(j)}} = \frac{1}{r\xi^{(j)}}\frac{d(\bullet)}{d\overline{s}^{(j)}}, \tag{B1}$$

where the subscript $i$ = 1, 2, 3 represents the component of the quantities along either the local ($\mathbf{e}_i$) or global ($\mathbf{E}_i$) coordinate directions, the superscript $j$ denotes the quantities associated with the $j$-th segment, $r$ is the corner radius, and $\xi^{(j)} = l^{(j)} / r$ is a scaling factor to unify the normalized arc length $\overline{s}^{(j)}$ of each segment to the same interval [0, 1] with $l^{(j)}$ being the length of the $j$-th segment. The lengths of the straight edges and rounded corners in the square rings are $L-2r$ and $\pi r/2$, respectively, where $L$ is the edge length of the corresponding square ring without rounded corners. For hexagonal rings, the corresponding lengths are $L-2r/\sqrt{3}$ for the straight edges and $\pi r/3$ for the corners. Additionally, in type-II square and hexagonal rings, the natural curvature undergoes a jump at the two ends of each corner segment (marked as pink in the schematics in **Fig. B1**), either from $\kappa_\mathrm{n}$ to $-\kappa_\mathrm{n}$ or vice versa, where $\kappa_\mathrm{n}$ denotes the magnitude of the natural curvature along the straight edges. To ensure continuity of natural curvature throughout the ring, we assume a linear variation of natural curvature across the rounded corners. For instance, in segment ② of the type-II square ring, the dimensionless natural curvature is given by $\overline{\kappa}_\mathrm{n}^{(2)} = 2\overline{\kappa}_\mathrm{n}\overline{s}^{(2)} - \overline{\kappa}_\mathrm{n}$, while in segment ④ of the type-II hexagonal ring, it is $\overline{\kappa}_\mathrm{n}^{(4)} = -2\overline{\kappa}_\mathrm{n}\overline{s}^{(4)} + \overline{\kappa}_\mathrm{n}$.

Using Eq. (B1), the dimensionless governing equations can be obtained as

$$\begin{aligned}
&\overline{N}_1^{(j)\prime} = \left(\overline{N}_2^{(j)}\overline{\kappa}_3^{(j)} - \overline{N}_3^{(j)}\overline{\kappa}_2^{(j)}\right)\xi^{(j)}, \overline{N}_2^{(j)\prime} = \left(\overline{N}_3^{(j)}\overline{\kappa}_1^{(j)} - \overline{N}_1^{(j)}\overline{\kappa}_3^{(j)}\right)\xi^{(j)}, \overline{N}_3^{(j)\prime} = \left(\overline{N}_1^{(j)}\overline{\kappa}_2^{(j)} - \overline{N}_2^{(j)}\overline{\kappa}_1^{(j)}\right)\xi^{(j)},\\
&\overline{\kappa}_1^{(j)\prime} = \overline{\kappa}_0^{(j)\prime} + \left[\overline{N}_2^{(j)} + (\beta-1)\overline{\kappa}_2^{(j)}\overline{\kappa}_3^{(j)} - \beta\overline{\kappa}_\mathrm{n}^{(j)}\overline{\kappa}_3^{(j)}\right]\xi^{(j)} / \alpha,\\
&\overline{\kappa}_2^{(j)\prime} = \overline{\kappa}_\mathrm{n}^{(j)\prime} - \left[\overline{N}_1^{(j)} + (\alpha-1)\overline{\kappa}_1^{(j)}\overline{\kappa}_3^{(j)} - \alpha\overline{\kappa}_0^{(j)}\overline{\kappa}_3^{(j)}\right]\xi^{(j)} / \beta,\\
&\overline{\kappa}_3^{(j)\prime} = \left[(\alpha-\beta)\overline{\kappa}_1^{(j)}\overline{\kappa}_2^{(j)} - \alpha\overline{\kappa}_0^{(j)}\overline{\kappa}_2^{(j)} + \beta\overline{\kappa}_\mathrm{n}^{(j)}\overline{\kappa}_1^{(j)}\right]\xi^{(j)},\\
&\overline{p}_1^{(j)\prime} = 2\left(q_1^{(j)}q_3^{(j)} + q_0^{(j)}q_2^{(j)}\right)\xi^{(j)}, \overline{p}_2^{(j)\prime} = 2\left(q_2^{(j)}q_3^{(j)} - q_0^{(j)}q_1^{(j)}\right)\xi^{(j)}, \overline{p}_3^{(j)\prime} = 2\left[\left(q_0^{(j)}\right)^2 + \left(q_3^{(j)}\right)^2 - 0.5\right]\xi^{(j)},\\
&q_0^{(j)\prime} = \left(-q_1^{(j)}\overline{\kappa}_1^{(j)} - q_2^{(j)}\overline{\kappa}_2^{(j)} - q_3^{(j)}\overline{\kappa}_3^{(j)}\right)\xi^{(j)} / 2, q_1^{(j)\prime} = \left(q_0^{(j)}\overline{\kappa}_1^{(j)} - q_3^{(j)}\overline{\kappa}_2^{(j)} + q_2^{(j)}\overline{\kappa}_3^{(j)}\right)\xi^{(j)} / 2,\\
&q_2^{(j)\prime} = \left(q_3^{(j)}\overline{\kappa}_1^{(j)} + q_0^{(j)}\overline{\kappa}_2^{(j)} - q_1^{(j)}\overline{\kappa}_3^{(j)}\right)\xi^{(j)} / 2, q_3^{(j)\prime} = \left(-q_2^{(j)}\overline{\kappa}_1^{(j)} + q_1^{(j)}\overline{\kappa}_2^{(j)} + q_0^{(j)}\overline{\kappa}_3^{(j)}\right)\xi^{(j)} / 2.
\end{aligned} \tag{B2}$$

*B.2. Boundary conditions*

In this work, we solve two types of BVPs: (i) the static equilibrium of a planar ring with out-of-plane natural curvature, and (ii) the transition behavior of a planar ring with out-of-plane natural curvature under external bending loads. In both cases, the governing equations for the rod segments are the same, i.e., Eq. (B2). Next, we introduce the corresponding boundary conditions for different types of rings.

***Square rings and type-I hexagonal rings***

For type-I and type-II square rings and type-I hexagonal rings, only one quarter of the ring is used in the analysis. The boundary conditions for analyzing their transition behavior under external bending loads are identical to those derived previously for curved-sided hexagram rings (Lu et al., 2023a). At the left end of the first segment (i.e., the midpoint of the bottom edge of the square and hexagonal rings shown in **Fig. B1(a-c)**), the boundary conditions are given by

$$\begin{aligned}&\bar{N}_2^{(1)}(0)=0,\bar{p}_1^{(1)}(0)=0,\bar{p}_3^{(1)}(0)=0,\\&\bar{\kappa}_3^{(1)}(0)=0,q_1^{(1)}(0)=0,q_2^{(1)}(0)=0.\end{aligned}\tag{B3}$$

At the right end of the $m$-th segment ($m$ is the total segment number of the quarter ring), where the bending load is applied, the boundary conditions are written as

$$\begin{aligned}&\bar{N}_1^{(m)}(1)=0,\bar{N}_2^{(m)}(1)=0,\bar{p}_2^{(m)}(1)=0,\\&q_0^{(m)}(1)=\frac{\sqrt{2}}{2}\cos\left(\frac{\Theta}{2}\right),q_1^{(m)}(1)=\frac{\sqrt{2}}{2}\cos\left(\frac{\Theta}{2}\right),\\&q_2^{(m)}(1)=\frac{\sqrt{2}}{2}\sin\left(\frac{\Theta}{2}\right),q_3^{(m)}(1)=-\frac{\sqrt{2}}{2}\sin\left(\frac{\Theta}{2}\right).\end{aligned}\tag{B4}$$

In Eq. (B4), $\Theta$ is the prescribed bending angle, which is applied with respect to the $\mathbf{e}_3$-direction, as illustrated in **Fig. B2**. Note that the bending angle is related to the quaternions through Euler angles. At the right loading point, the Euler angles ($\varphi$, $\theta$, $\phi$), following the 3-2-3 rotation convention that relates the local basis ($\mathbf{e}_1$, $\mathbf{e}_2$, $\mathbf{e}_3$) to the global basis ($\mathbf{E}_1$, $\mathbf{E}_2$, $\mathbf{E}_3$), are given by ($-\pi/2$, $\pi/2$, $\pi/2-\Theta$). Using the relationship between Euler angles and quaternions (Healey and Mehta, 2005; Yu and Hanna, 2019),

$$q_0 = \cos\frac{\theta}{2}\cos\frac{\phi+\varphi}{2}, q_1 = \sin\frac{\theta}{2}\sin\frac{\phi-\varphi}{2}, q_2 = \sin\frac{\theta}{2}\cos\frac{\phi-\varphi}{2}, q_3 = \cos\frac{\theta}{2}\sin\frac{\phi+\varphi}{2}, \tag{B5}$$

the boundary conditions associated with quaternions in Eq. (B4) are obtained.

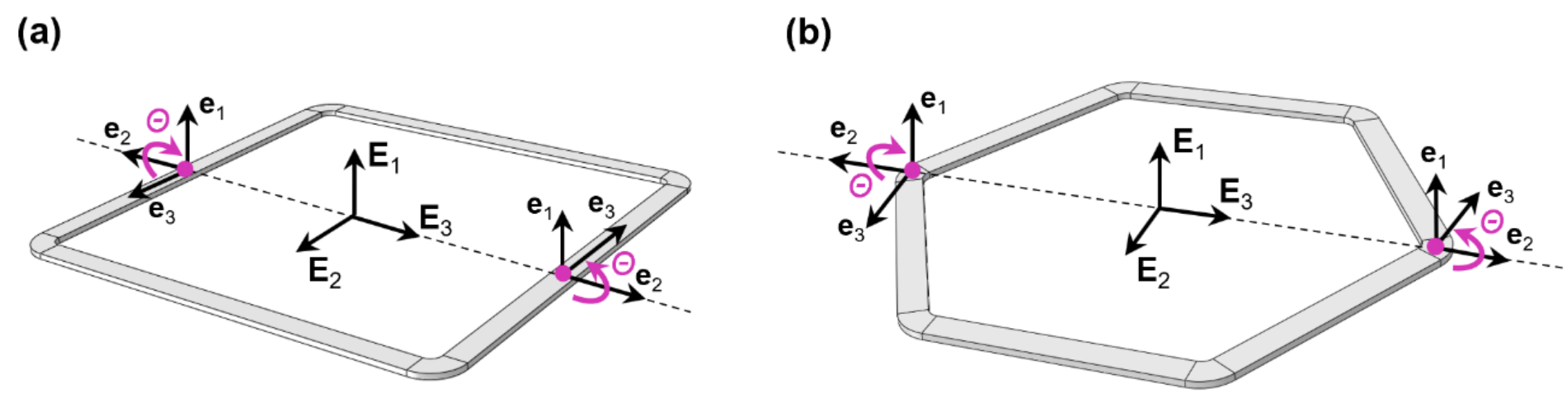

**Fig. B2.** Schematics of (a) square and (b) hexagonal rings subjected to a pair of bending loads. The pink dots represent the loading positions. The bending angle Θ is applied with respect to the $\mathbf{e}_3$-direction.

Based on our experimental observations and FEA simulations, the out-of-plane natural curvature induces an internal bending moment that drives the ring edge to deform out of plane, analogous to deformation caused by external bending loads. As a result, the boundary conditions used to determine the equilibrium states of square and hexagonal rings with out-of-plane natural curvature are similar to those used in the analysis of transitions under external bending loads, i.e., Eqs. (B3) and (B4). In particular, the left end of the first segment is free of external loads in both cases, so the boundary conditions are identical, i.e., Eq. (B3). At the right end of the *m*-th segment, no bending angle is prescribed. Therefore, the boundary conditions associated with the quaternions in Eq. (B4) can be written in its equivalent form as $q_0^{(m)} = q_1^{(m)}$ and $q_2^{(m)} = -q_3^{(m)}$. The quaternion normalization condition then becomes $\left(q_0^{(m)}\right)^2 + \left(q_2^{(m)}\right)^2 = 0.5$. Furthermore, since the *m*-th segment is free to rotate about the $\mathbf{e}_3$-direction at the right end, the associated twisting moment vanishes, implying $\bar{\kappa}_3^{(m)} = 0$. Consequently, the boundary conditions at the right end of the *m*-th segment to determine the equilibrium states are given by

$$\begin{aligned}
&\bar{N}_1^{(m)}(1) = 0, \bar{N}_2^{(m)}(1) = 0, \bar{p}_2^{(m)}(1) = 0, \bar{\kappa}_3^{(m)}(1) = 0, \\
&q_0^{(m)}(1) - q_1^{(m)}(1) = 0, q_2^{(m)}(1) + q_3^{(m)}(1) = 0, \\
&\left[q_0^{(m)}(1)\right]^2 + \left[q_2^{(m)}(1)\right]^2 - 0.5 = 0.
\end{aligned} \tag{B6}$$

### *Type-II hexagonal rings*

For type-II hexagonal rings, half of the ring is used in the analysis, as shown in **Fig. B1(d)**. When a pair of bending loads are applied at the left end of the first segment and right end of the $m$-th segment, the boundary conditions at the right end are identical to those in Eq. (B4). At the left end, the translational degrees of freedom (DOFs) are constrained to eliminate rigid-body motion, resulting in $\bar{p}_1^{(1)} = 0$, $\bar{p}_2^{(1)} = 0$, and $\bar{p}_3^{(1)} = -\sqrt{3}L/(2r)$. For the rotational DOFs, constraints are first defined using Euler angles and then converted into quaternions. The Euler angles at the left end of the first segment for the type-II hexagonal ring are $(\varphi, \theta, \phi) = (-\pi/2, -\pi/2, \pi/2-\Theta)$. Substituting it into Eq. (B5), the corresponding quaternions are obtained as

$$
\begin{aligned}
&q_0^{(1)}(0) = \frac{\sqrt{2}}{2}\cos\left(\frac{\Theta}{2}\right), q_1^{(1)}(0) = -\frac{\sqrt{2}}{2}\cos\left(\frac{\Theta}{2}\right), \\
&q_2^{(1)}(0) = -\frac{\sqrt{2}}{2}\sin\left(\frac{\Theta}{2}\right), q_3^{(1)}(0) = -\frac{\sqrt{2}}{2}\sin\left(\frac{\Theta}{2}\right).
\end{aligned} \tag{B7}
$$

The left and right ends together impose fourteen boundary conditions, which is one more than required. To avoid over-constraint, only three of the quaternions at the left end are specified (the remaining one is self-satisfied due to the quaternion normalization condition). Accordingly, the six boundary conditions at the left end of the first segment for analyzing the transition behavior under bending loads are:

$$
\begin{aligned}
&\bar{p}_1^{(1)}(0) = 0, \bar{p}_2^{(1)}(0) = 0, \bar{p}_3^{(1)}(0) = -\sqrt{3}L/(2r), \\
&q_0^{(1)}(0) = \frac{\sqrt{2}}{2}\cos\left(\frac{\Theta}{2}\right), q_2^{(1)}(0) = -\frac{\sqrt{2}}{2}\sin\left(\frac{\Theta}{2}\right), q_3^{(1)}(0) = -\frac{\sqrt{2}}{2}\sin\left(\frac{\Theta}{2}\right).
\end{aligned} \tag{B8}
$$

Similar to the case of square rings, the boundary conditions at the left end of the first segment for determining the equilibrium states of type-II hexagonal rings can be expressed as

$$
\begin{aligned}
&\bar{p}_1^{(1)}(0) = 0, \bar{p}_2^{(1)}(0) = 0, \bar{p}_3^{(1)}(0) = -\sqrt{3}L/(2r), \\
&\bar{\kappa}_3^{(1)}(0) = 0, q_0^{(1)}(0) + q_1^{(1)}(0) = 0, q_2^{(1)}(0) - q_3^{(1)}(0) = 0.
\end{aligned} \tag{B9}
$$

At the right end of the $m$-th segment, the boundary conditions are identical to those given in Eq. (B6), except that $\bar{N}_1^{(m)}(1) = 0$ is replaced by $\bar{p}_1^{(m)}(1) = 0$ to avoid rigid-body motion.

Apart from the thirteen boundary conditions imposed at the left end of the first segment and the right end of the *m*-th segment, the remaining 13(*m*-1) boundary conditions at the interfaces between adjacent segments can be readily obtained by enforcing the force equilibrium and geometric compatibility (Lu et al., 2023a). These 13(*m*-1) conditions are applied consistently for analyzing both the equilibrium states and transition behavior. Once the multi-segment rod system is solved, the total strain energy of the ring can be calculated by

$$U = \frac{\eta}{2}\sum_{j=1}^{m}\left[\int EI_1\left(\kappa_1^{(j)} - \kappa_0^{(j)}\right)^2 ds^{(j)} + \int EI_2\left(\kappa_2^{(j)} - \kappa_{\mathrm{n}}^{(j)}\right)^2 ds^{(j)} + \int GJ\left(\kappa_3^{(j)}\right)^2 ds^{(j)}\right], \quad \text{(B10)}$$

where $\eta = 4$ for type-I and type-II square rings and type-I hexagonal rings, and $\eta = 2$ for type-II hexagonal rings.

## Appendix C. Finite element simulations

Finite element simulations for the equilibrium states and transition behavior of elastic rings with out-of-plane natural curvature are performed in the commercial software Abaqus 2024 (Dassault Systèmes, France). The ring is simulated using a 3D shell model, and the S4R (4-node stress/displacement shell element with reduced integration) linear element is employed. In all simulations, the height and width of the ring's cross-section are taken as $h = 0.5$ mm and $t = 2$ mm, respectively, and the radius of the rounded corner is considered as $r = 2$ mm. For square rings, the edge length is taken as $L = 200$ mm, and for hexagonal rings, the edge length is taken as $L = 100$ mm. Moreover, the Young's modulus and Poisson's ratio of the ring are taken as $E = 200$ GPa and 0.3, respectively.

To simulate the equilibrium state of the ring, a temperature gradient is prescribed across its height direction, inducing a linear strain distribution analogous to that caused by natural curvature. The magnitude of the temperature gradient $\nabla T$ is determined by equating the resultant thermal moment $EI_2 \nabla T \alpha_{\mathrm{T}}$ to the natural curvature-induced bending moment $-EI_2\kappa_{\mathrm{n}}$, leading to $\nabla T = -\kappa_{\mathrm{n}} / \alpha_{\mathrm{T}}$, where $\alpha_{\mathrm{T}} = 0.002\ \mathrm{K}^{-1}$ is the specified coefficient of thermal expansion (CTE) along the longitudinal direction of the rod segment (note that the thermal expansion is defined as orthotropic in the simulation, and the CTEs in the other two directions are set to zero). To simulate the transition behavior of the ring under external bending loads, the temperature gradient is first applied to replicate the effect of natural curvature, and then a pair of rotational angles are imposed at the loading points to trigger the transition. Since the rotational angles at the loading points are initially nonzero in the 3D equilibrium configuration, they are first reset to zero to capture the complete transition process with rotational angles varying from 0 to $2\pi$. Additionally, a small damping factor of $10^{-8}$ is introduced to stabilize the simulation.